\documentclass[iicol, pdflatex,sn-mathphys-num]{sn-jnl}

\usepackage{graphicx}%
\usepackage{multirow}%
\usepackage{amsmath,amssymb,amsfonts}%
\usepackage{amsthm}%
\usepackage{mathrsfs}%
\usepackage[title]{appendix}%
\usepackage{xcolor}%
\usepackage{textcomp}%
\usepackage{manyfoot}%
\usepackage{booktabs}%
\usepackage{algorithm}%
\usepackage{algorithmicx}%
\usepackage{algpseudocode}%
\usepackage{listings}%

\usepackage{makecell}%
\usepackage{threeparttable}%
\usepackage{colortbl}
\usepackage{booktabs}
\usepackage{placeins}
\usepackage{xurl}
\mathchardef\UrlBreakPenalty=5000
\mathchardef\UrlBigBreakPenalty=200

\newcommand{\rev}[1]{\textcolor{black}{#1}}
\usepackage[switch,mathlines]{lineno}

\definecolor{paperbg}{RGB}{247,246,242}
\definecolor{headbg}{RGB}{244,244,240}
\definecolor{labelbg}{RGB}{249,249,246}
\definecolor{textdark}{RGB}{45,45,45}

\definecolor{fg1}{RGB}{236,244,238}
\definecolor{fg2}{RGB}{214,231,219}
\definecolor{fg3}{RGB}{169,197,178}
\definecolor{fg4}{RGB}{86,118,101}

\definecolor{pp1}{RGB}{239,235,246}
\definecolor{pp2}{RGB}{219,210,235}
\definecolor{pp3}{RGB}{176,157,207}
\definecolor{pp4}{RGB}{106,84,145}

\definecolor{gd1}{RGB}{248,242,223}
\definecolor{gd2}{RGB}{234,220,176}
\definecolor{gd3}{RGB}{208,177,96}
\definecolor{gd4}{RGB}{170,132,44}

\theoremstyle{thmstyleone}%
\theoremstyle{thmstyletwo}%

\theoremstyle{thmstylethree}%

\begin{document}

\title[Article Title]{Diagnosing the Limits of Large Language Models for Structured Scientific Evidence Extraction}


\author*[1]{\fnm{Zhiyin} \sur{Tan}}\email{zhiyin.tan@l3s.de}

\author[2]{\fnm{Jennifer} \sur{D’Souza}}\email{jennifer.dsouza@tib.eu}

\author[3]{\fnm{Sören} \sur{Auer}}\email{auer@tib.eu}

\affil[1,3]{\orgdiv{L3S Research Center}, \orgname{Leibniz University Hannover}, \city{Hannover}, \country{Germany}}

\affil[2,3]{\orgname{TIB, Leibniz Information Centre for Science and Technology}, \city{Hannover}, \country{Germany}}


\abstract{
Scientific evidence is predominantly communicated through unstructured articles, while many downstream workflows require structured representations of study populations, variables, methods, and quantitative findings. Large language models (LLMs) offer a promising route toward automating this transformation, yet their reliability in constructing schema-constrained evidence records from parsed scientific full texts remains insufficiently understood. Evidence synthesis is a representative use case in which systematic reviews and meta-analyses require accurate, structured study information before quantitative integration can begin.
We present a schema-grounded, tiered evaluation framework for structured scientific evidence extraction that increases task difficulty from single-entity identification to relational tuples and corpus-level statistical tasks. We evaluate frontier commercial and open-weight models across five scientific domains under a stricter multi-level assessment protocol, complemented by human validation of model-based judging and process-level inspection of intermediate reasoning for statistical tasks.
Results reveal a clear pattern: models perform substantially better on low-complexity extraction tasks than on tasks requiring correct bindings among entities, numerical values, and methodological attributes. Intermediate-step analysis further indicates error amplification under exact downstream aggregation, where small upstream errors often prevent correct final outputs. Scaling model size yields uneven gains and does not remove these weaknesses.
These findings show that recent advances have not resolved the fundamental challenges of structured scientific evidence extraction. We provide a reusable query suite and diagnostic evaluation framework to support future fine-grained benchmarks and more rigorous assessment of models for evidence-intensive scientific workflows.
}




\maketitle

\section{Introduction}

Scientific knowledge is predominantly communicated through unstructured articles, yet many downstream workflows depend on structured representations of study populations, variables, methods, and quantitative findings. Researchers conducting evidence synthesis must transform narrative study reports into tabular records before quantitative integration can begin \cite{cochrane2019handbook,cooper2015research}. Scholarly knowledge platforms increasingly rely on explicit entities and relations to support comparison, retrieval, and reuse \cite{auer2025open,ammar2018S2LiteratureGraph,dessi2025cskg}. Domain databases and monitoring systems likewise depend on normalized study descriptors and results that can be queried computationally. Across these settings, a recurring bottleneck is converting free-form scientific prose into reliable, structured evidence records.
This transformation remains highly labor-intensive. In systematic reviews and related workflows, extracting study-level information such as populations, variables, sample sizes, measurement units, statistical methods, and reported effects often requires careful manual reading and verification across many articles \cite{tsafnat2014systematic}. Beyond evidence synthesis, building structured scientific resources typically depends on expensive annotation pipelines, handcrafted rules, or narrowly scoped extraction systems. As scientific publishing continues to grow, scalable methods for structured extraction of scientific evidence are increasingly needed.
Large language models (LLMs) offer a promising route toward this goal. Recent systems have shown strong general-purpose language capabilities and encouraging results in scientific text mining, information extraction, and structured generation \cite{shamsabadi2024large,dagdelen2024structured,lu2025finetuning,d2025mining}. However, current capabilities remain insufficiently understood for the forms of extraction that matter most in practice.

To study this systematically, we evaluate models under a tiered framework that increases structural difficulty from single-field extraction to bound tuples to corpus-level statistical tasks over extracted evidence. Many existing evaluations focus on isolated entities, short-context question answering, or domain-specific benchmarks with relatively local targets. Yet the extraction of real scientific evidence often requires assembling multiple interdependent fields into coherent, structured records. Limitations in compositionality, entity binding, long-context evidence use, and numerical grounding are therefore especially relevant, but are rarely examined within one unified setting \cite{Dziri2023Neural,ismayilzada2025LLMMorphological,feng2024LLMBindEntities,gurarieh2025LLMRetrieveBoundEntities,liu2024LLMLostInLongContext,akhtar2023LLMNumericalReasoningCapabilities,ahmed2025LLMNumericalReasoningCapabilities,oyesanmi2025LLMReasoning}.
Evidence synthesis is one representative real-world example. Systematic reviews and meta-analyses depend on extracting structured study information before results can be compared or quantitatively integrated. However, the same requirements also arise in scientific databases, scholarly knowledge graphs, automated literature analysis, and cross-paper comparison systems. We therefore treat evidence synthesis as one visible use case of structured scientific evidence extraction rather than the sole target application.

In this work, we present a schema-grounded, tiered evaluation framework for structured scientific evidence extraction. We study both task complexity and input organization by comparing a Global long-context regime with a Per-paper decomposed regime. Queries progress from single-entity identification to higher-arity relational tuples and corpus-level statistical tasks over extracted evidence structures.
This paper substantially extends our earlier conference paper~\cite{TD2026meta}\rev{, presented at the Conference on Information and Research Science Connecting to Digital and Library Science (IRCDL 2026). As the proceedings have not yet been published, we cite the publicly available preprint version}. While retaining the same five-domain corpus and overall problem setting, we re-evaluate the framework using updated frontier commercial and open-weight models across multiple model scales, replace the original evaluation setup with a stricter multi-level assessment protocol, add human validation of model-based judging, introduce process-level inspection of intermediate reasoning steps for statistical tasks, and refine portions of the query suite. These extensions enable a more robust and current assessment of whether recent model advances have resolved previously observed limitations.
Our empirical results show that performance declines sharply as structural demands increase\rev{, that is, as tasks move from recovering a single isolated field, to jointly extracting several interdependent fields (e.g., variable pairs, the statistical method, and the effect size) that must be correctly bound within one coherent record, and finally to computing exact corpus-level statistics over such records}. Models are stronger on simpler extraction tasks, but weaker on multi-field binding and exact downstream outputs. Process-level analysis suggests evidence of error amplification under exact downstream aggregation, where partial evidence recovery often fails when strict exactness is required.
This paper makes \textbf{three main contributions}:
\begin{enumerate}
    \item we introduce a schema-grounded framework for evaluating structured scientific evidence extraction through progressively harder query types;
    \item we provide a reusable cross-domain query suite that supports future fine-grained benchmarks and comparative studies;
    \item we deliver an updated empirical diagnosis of frontier model behavior on full-text scientific extraction tasks, identifying persistent challenges that future systems must address.
\end{enumerate}
Together, these results clarify the current limits of automated scientific evidence structuring and motivate more rigorous next-generation evaluations.
The released resources are described in the Data availability and Code availability statements.

\section{Related Work}

\paragraph{Structured Scientific Information Extraction}

A substantial body of research in semantic publishing and scholarly knowledge representation has shown that scientific outputs can be modeled through explicit schemas, ontologies, and knowledge graphs. Prior work formalizes article structure in machine-interpretable form~\cite{Constantin2016DoCO}, represents scholarly entities and citation relations through interoperable ontology modules~\cite{Peroni2012FaBiOandCiTO,peroni2018spar}, and, particularly in biomedicine, captures hypotheses, claims, and evidence as structured discourse objects rather than only prose~\cite{Ciccarese2008SWANbiomedicalDiscourseOntology}. Large-scale scientific knowledge graphs further demonstrate that complex research entities and relations can be integrated into reusable infrastructures, including biomedical resources such as SemMedDB~\cite{Kilicoglu2012SemMedDB}, integrative graph systems such as Hetionet~\cite{Himmelstein2017SystematicIntegrationBiomedical}, and domain-specific scientific graphs such as MatKG~\cite{venugopal2024matkg}.
Within this landscape, the Open Research Knowledge Graph (ORKG) provides infrastructure for authoring structured paper-level contribution descriptions and comparing findings across publications~\cite{oelen2019ORKGcomparing,auer2025open}. These efforts collectively support the view that scientific studies can be represented as structured objects whose key components, such as populations, variables, methods, and outcomes, are amenable to schema-guided information extraction. However, much of this literature focuses on representation design, curation workflows, or downstream infrastructure rather than systematically evaluating whether current frontier language models can reliably populate such structures from full scientific texts.
Parallel work in scientific information extraction has developed methods for recovering entities and relations directly from research articles. Document-level benchmarks such as SciREX emphasize that important scientific entities and relations are often distributed across sections and must be assembled into coherent, structured records~\cite{jain2020SciREX}. 

More recent datasets and evaluations likewise show that extracting fine-grained full-text relations remains challenging and often requires more than local span prediction~\cite{zhang2024sciER,pham2023LabelVariationScientificInfoExtraction}. Multi-document and multimodal scientific reasoning benchmarks further report substantial gaps between current systems and expert performance when evidence must be integrated across papers or modalities~\cite{li2024M3SciQA,park2025Multi-docMulti-hopQA,T2025MD}. 

Our work builds on this line of research but shifts the focus from task-specific extraction accuracy to a broader diagnostic question: how performance changes as the demands of structured extraction become progressively more complex.
Concretely, we operationalize scientific evidence as study-level tuples over recurring schema fields such as populations, variables, methods, and quantitative findings, which form the basis of the evaluation framework introduced next.

\begin{figure*}[!htb]
  \centering
  \includegraphics[width=\linewidth]{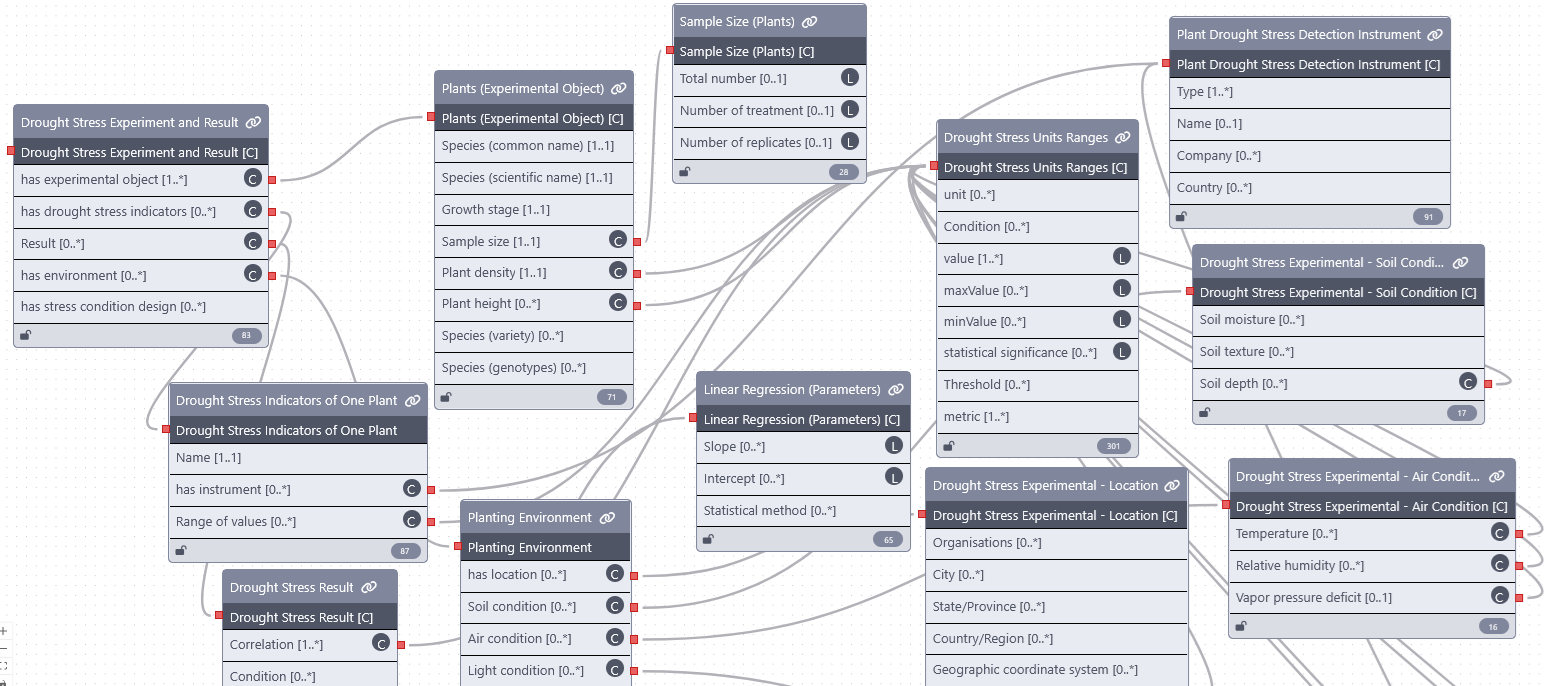}
  \caption{Partial view of an earlier ORKG template for \textit{Early Drought Stress Indicators in Plants (Physiological \& Multi-Sensor)} \url{https://orkg.org/templates/R1471884}. The original template contains richly nested domain-specific fields for plant descriptors, sensor indicators, environmental conditions, and quantitative outcomes. In this study, we abstract recurrent study-level components from such domain templates into a simplified cross-domain schema for structured scientific evidence extraction.}
  \label{fig:orkg-semantic-model}
\end{figure*}

\paragraph{LLMs in Evidence-Intensive Scientific Workflows}

A growing literature examines the use of language models in workflows that depend on scientific evidence extraction and synthesis. In systematic reviews and meta-analyses, information extraction, e.g., data or evidence abstraction, is a central step in which reviewers convert study reports into structured records~\cite{cochrane2019handbook,Schmidt2025DataExtractionforReview}. Prior work has also documented the difficulty and error-proneness of manual extraction procedures, including higher error rates under single-extractor workflows~\cite{Buscemi2006SingleDataExtractionReviews}.
Automation research spans earlier rule-based and statistical approaches for extracting PICO-style elements from biomedical texts~\cite{Demner-Fushman2007ClinicalQA-KG-Statistical,Wallace2016PICO}, systems targeting structured trial characteristics \cite{Kiritchenko2010ExaCT}, and newer neural approaches for more precise extraction of study attributes and relations \cite{Hu2023PICOextraction}. More recent studies explore language models as assistants across multiple stages of the review pipeline, including screening, bias assessment, and study-level extraction~\cite{Lieberum2025LLMNotReadyforReview,Khan2025CollaborativeLLMsDataExtractionReview,Brassey2021AutomatedEvidenceSynthesis,Jardim2022RiskOfBiasAutomation}.
Beyond evidence synthesis, recent work applies language models to scientific information extraction tasks such as schema-guided entity recovery, structured generation, and extraction of n-ary scientific records from full papers~\cite{dagdelen2024structured,d2025mining,kabongo2024effective}. Related research in multi-document question answering studies how models gather and combine evidence across documents, often through structured intermediates such as knowledge graphs~\cite{Wang2024KGMulti-DocumentQA,yang2025LLMMulti-DocQAKGReasoning}. Collectively, these studies demonstrate growing interest in evidence-intensive scientific workflows. 
However, many evaluations focus on end-task outputs or narrowly scoped extraction targets, leaving the reliability of schema-constrained, multi-field scientific evidence records less well understood. To study this directly, we next define a compact study-level schema covering recurring entities, relations, and reported quantitative results across domains.


\begin{table*}[!htb]
\caption{Study-level schema properties used for structured scientific evidence extraction.}
\label{tab:schema}
\centering
\small
\begin{tabular}{p{3.8cm} p{8.5cm} p{2.2cm}}
\toprule
\textbf{Property} & \textbf{Definition} & \textbf{Notation} \\
\midrule
Document ID & Index of the source document in the corpus & $i \in \{1,\ldots,|\mathcal{D}|\}$ \\

\midrule
Population & Study population or unit of analysis & $P \in \mathcal{P}$ \\
Geolocation & Country or region of study conduct & $G \in \mathcal{G}$ \\
Sample Size & Main reported sample size & $N \in \mathcal{N}$ \\

\midrule
Independent Variable & Predictor or explanatory variable & $IV \in \mathcal{IV}$ \\
Independent Variable Unit & Unit of measurement for IV & $U_{IV} \in \mathcal{U}_{IV}$ \\

Dependent Variable & Outcome or response variable & $DV \in \mathcal{DV}$ \\
Dependent Variable Unit & Unit of measurement for DV & $U_{DV} \in \mathcal{U}_{DV}$ \\

Variable & Role-agnostic variable (de-duplicated the union of $IV$ \& $DV$) & $V \in \mathcal{V}$ \\
Variable Unit & Unit of measurement for Variable & $U \in \mathcal{U}$ \\
\midrule
Association Method & Statistical model for association & $A \in \mathcal{A}$ \\
Conditions & Optional contextual qualifier for an association estimate & $C \in \mathcal{C}$ \\
Effect Size & Reported association estimate & $E \in \mathcal{E}$ \\
\bottomrule
\end{tabular}
\end{table*}


\paragraph{Model Limitations Relevant to Structured Evidence Extraction}

A separate line of work documents broader limitations of modern language models that are highly relevant to structured scientific extraction. Studies on compositional generalization show that models often struggle to combine familiar components into systematically correct novel structures~\cite{Dziri2023Neural,zhao2024CompositionalDeficiencyLLMMathematicalReasoning,Shojaee2025IllusionOfThinking}. Other work reports failures in reasoning fidelity and intermediate consistency, where fluent outputs may mask unstable internal reasoning processes~\cite{song2025LLMReasoningFailures,Mahowald2024LLMLanguageAndThought}.
Entity and attribute binding pose an additional challenge. Prior studies show that models may retrieve or recognize relevant elements while attaching them to incorrect counterparts~\cite{feng2024LLMBindEntities,gurarieh2025LLMRetrieveBoundEntities}. Long-context evaluations similarly report degradation when relevant evidence is distributed across extended inputs, including reduced attention to information located in the middle of long contexts~\cite{liu2024LLMLostInLongContext}. Numerical reasoning introduces further difficulties, especially when values must be compared, computed, or correctly grounded to specific entities and variables~\cite{akhtar2023LLMNumericalReasoningCapabilities,ahmed2025LLMNumericalReasoningCapabilities,oyesanmi2025LLMReasoning,boye2025LLMMathematicalReasoningFailures,yan2025LLMElementaryAddition}.
These limitations have usually been studied in isolation or in generic reasoning settings. Our work examines them within a shared schema of study-level evidence records constructed from full research articles across domains, which we define in the next section.

\section{Methodology}
\label{sec:method}

\subsection{Semantic Study Schema and Scope}

The schema used in this study builds on earlier semantic modeling work in the Open Research Knowledge Graph (ORKG), tailored to plant drought-stress research, as shown in Figure~\ref{fig:orkg-semantic-model}. Rather than retaining domain-specific nested fields, we abstract recurring study components into a compact cross-domain schema for evaluation.
In the present work, we generalize and simplify these earlier templates by retaining study elements that recur across empirical quantitative research. The resulting schema captures populations, geolocations, sample sizes, variables and units, association methods, and reported quantitative findings.
These fields are relevant across multiple downstream settings, including evidence synthesis, cross-paper comparison, structured indexing, and scientific knowledge management. We therefore use them as a common schema for evaluating structured evidence extraction across scientific domains.
Given this generalized schema, we focus on empirical studies that report statistical associations between variables, including correlation coefficients, standardized regression coefficients, coefficients of determination, and odds ratios~\cite{cochrane2019handbook,borenstein2021IntroMetaAnalysis}. Such studies are common across the behavioral, social, medical, environmental, and agricultural sciences, and they provide a suitable testbed for structured evidence extraction. While reporting conventions differ by discipline, recovering complete study findings still requires accurate entity binding, numerical grounding, and disambiguation of many-to-many relationships among variables across studies.

Table~\ref{tab:schema} summarizes the study-level schema $\mathcal{S}$ used in our evaluation.

\begin{table*}[!htb]
\centering
\caption{Object-centric (O) list-style extraction queries. All outputs preserve the source document identifier $i$.}
\label{tab:object-queries}
\small
\begin{tabular}{p{1.4cm} p{2.7cm} p{10.5cm}}
\toprule
\textbf{ID} & \textbf{Tuple Pattern} & \textbf{Natural-Language Query} \\
\midrule
O1.1 & $(i, G)$       & Extract the study geolocation (no inference). \\
O1.2 & $(i, N)$       & Extract the final analytic sample size (\rev{\texttt{`not reported'}} if unavailable). \\
O1.3 & $(i, P)$       & Extract the study population/unit of analysis. \\
\midrule
O2.1 & $(i, P, N)$    & Extract each study population paired with its sample size. \\
O2.2 & $(i, P, G)$    & Extract each study population paired with its geolocation. \\
O2.3 & $(i, P, G, N)$ & Extract each study population with both geolocation and sample size. \\
\bottomrule
\end{tabular}
\end{table*}


\begin{table*}[!htb]
\centering
\caption{Method-centric (M) list-style extraction queries. All outputs preserve the source document identifier $i$.}
\label{tab:method-queries}
\small
\begin{tabular}{p{1.4cm} p{2.7cm} p{10.5cm}}
\toprule
\textbf{ID} & \textbf{Tuple Pattern} & \textbf{Natural-Language Query} \\
\midrule
M1.1 & $(i, A)$               & Extract the statistical method used to quantify associations. \\
M1.2 & $(i, V)$               & Extract all variables (union of IVs and DVs) appearing in association tests. \\
M1.3 & $(i, IV)$              & Extract all independent variables (predictor construct). \\
M1.4 & $(i, DV)$              & Extract all dependent variables (outcome construct). \\
\midrule
M2.1 & $(i, V, U)$         & Extract each variable together with its measurement unit. \\
M2.2 & $(i, IV, U_{IV})$        & Extract each independent variable with its measurement unit. \\
M2.3 & $(i, DV, U_{DV})$        & Extract each dependent variable with its measurement unit. \\
M2.4 & $(i, IV, DV)$          & Extract independent--dependent variable pairs. \\
M2.5 & $(i, IV, DV, A)$       & Extract variable pairs together with the statistical method. \\
M2.6 & $(i, IV, DV, A, C, E)$ & Extract variable pairs, statistical method, conditions and effect size. \\
\bottomrule
\end{tabular}
\end{table*}

\subsection{Problem Formulation}

Let $\mathcal{D} = \{d_1, \ldots, d_n\}$ be a corpus of empirical studies. For each study $d_i$, we conceptualize its extractable content as a set of tuples drawn from subsets of the schema properties:

\[
\mathcal{I}(d_i) = \{t_{i1}, t_{i2}, \ldots \}.
\]

Each tuple $t_{ij}$ is an ordered $k$-tuple $(X_1,\ldots,X_k)$, where each element corresponds to a schema field in $\mathcal{S}$. The tuple arity $k$ reflects how many fields must be jointly identified and correctly bound.

We distinguish two tasks:
\begin{enumerate}
    \item \textbf{Tiered structural extraction tasks}, which request all instances of a given tuple pattern across the corpus.
    \item \textbf{Corpus-level statistical tasks}, which operate over extracted evidence structures to produce lightweight summary outputs.
\end{enumerate}

In both cases, the input to the model is the document set together with a natural-language query. Outputs are constrained to predefined JSON formats. 
Our methodological objective is to examine how model behavior changes as extraction demands shift from local-entity recovery to higher-arity structured records and simple downstream summarization. We also test whether these demands are better handled by a single global context or by a per-paper, decomposed pipeline.

\paragraph{Task 1: Tiered Structural Extraction Tasks}
The structural component of the framework is divided into two complementary views. Object-centric queries focus on study descriptors such as population, sample size, and geolocation. Method-centric queries focus on variables, measurement units, analytical methods, and reported associations. Together, these two views cover both contextual metadata and substantive quantitative findings.

\bmhead{Object-centric queries}
The O-series tasks progress from single-field extraction to increasingly bound study descriptors. As shown in Table~\ref{tab:object-queries}, O1-level tasks request one attribute at a time, such as geolocation, final analytic sample size, or study population. O2-level tasks require joint recovery of multiple linked fields, for example pairing populations with sample size, geolocation, or both simultaneously. These tasks test whether models can preserve document-level study context while maintaining correct attribute associations.

\bmhead{Method-centric queries}
The M-series tasks examine the extraction of analytical content. As shown in Table~\ref{tab:method-queries}, M1-level tasks request individual fields such as the statistical method used, all variables appearing in association tests, or the independent and dependent variables separately. M2-level tasks require increasingly structured outputs, including variable-unit bindings, independent-dependent variable pairs, variable pairs linked to methods, and finally full association tuples containing variables, method, conditions, and effect size. These tasks place stronger demands on relational consistency and numerical grounding.

\bmhead{Progressive structural difficulty}
Across tasks, difficulty increases with tuple arity and dependency structure. Recovering isolated fields mainly requires recognition and normalization. Recovering higher-order tuples additionally requires that all fields originate from the correct study context and remain correctly bound within a coherent record.

\begin{table*}[!htb]
\centering
\caption{Derived statistical queries over document-attributed extraction results. Each query operates on specific schema atoms and applies a well-defined computation.}
\label{tab:computation-queries}
\small
\begin{tabular}{p{2cm} p{3.4cm} p{9.3cm}}
\toprule
\textbf{ID} & \textbf{Target Atoms} & \textbf{Computation} \\
\midrule
OC\_O1.1   & $(G \rightarrow \text{count})$           & Count unique countries explicitly stated across all documents. \\
OC\_O1.2.1 & $(N \rightarrow \text{count})$           & Count documents with $N > 100$. \\
OC\_O1.2.2 & $(N \rightarrow \text{mean})$            & Compute the arithmetic mean of $N$ across all documents. \\
OC\_O1.2.3 & $(N \rightarrow \text{median})$          & Compute the median of $N$ across all documents. \\
\midrule
MC\_M1.1 & $(A \rightarrow \text{count})$            & Count unique statistical methods across all documents. \\
MC\_M1.2 & $(V \rightarrow \text{count})$            & Count unique variables (union of IV and DV) across all documents. \\
MC\_M1.3 & $(IV \rightarrow \text{count})$           & Count unique independent variables across all documents. \\
MC\_M1.4 & $(DV \rightarrow \text{count})$           & Count unique dependent variables across all documents. \\
MC\_M2.4 & $((IV, DV) \rightarrow \text{count})$     & Count unique $(IV, DV)$ pairs across all documents. \\
MC\_M2.6 & $(IV, DV, E)$                              & List $(IV, DV)$ pairs with $|E| > 0.7$. \\
\bottomrule
\end{tabular}
\end{table*}

\paragraph{Task 2: Corpus-Level Statistical Tasks}

Beyond direct extraction, we define derived statistical queries that operate over document-attributed outputs. These tasks evaluate whether extracted structures remain sufficiently reliable for lightweight downstream use.
The object-centric computation tasks aggregate study descriptors, for example by counting explicitly stated countries, counting documents above a sample-size threshold, or computing mean and median sample size across documents. The method-centric computation tasks summarize analytical content, such as counting unique methods, variables, independent variables, dependent variables, or variable pairs, and identifying strong reported associations (Table~\ref{tab:computation-queries}).
These queries are intentionally simple in computational form. Their purpose is not to test advanced mathematical reasoning, but to examine whether errors introduced during extraction propagate into downstream summaries that depend on accurate structured inputs.

\bmhead{Structured Output Stages}
For computation-style queries, models must return a three-stage structured output. 
(i) Evidences, consists of document-attributed atoms extracted from individual papers and used as inputs to the computation. 
(ii) Final list, consolidates these paper-linked evidences through filtering, aggregation, or deduplication into an intermediate structured result; although merged at the corpus level, retained items remain attributable to contributing source papers. 
(iii) Final answer is the requested scalar statistic or final set-valued output and no longer carries paper-level identifiers.
This decomposition makes intermediate reasoning steps explicit and allows downstream analysis of whether failures arise from evidence collection, consolidation, or final answer generation.

\subsection{Evaluation Principles}
\label{sec:evaluation-principles}

We evaluate all outputs against manually curated gold annotations using tuple-level precision, recall, and F1. Because every prediction and gold annotation is explicitly linked to a source paper identifier, correctness requires not only semantic agreement on the extracted content, but also correct attribution to the originating paper.

\bmhead{Matching Procedure}
For each predicted tuple, candidate matches are restricted to gold tuples from the same source paper. This paper-level constraint ensures that extracted evidence is credited only when assigned to the correct document.
Within the eligible set, predicted and gold tuples are compared field by field under the three-level matching protocol defined in this subsection. For composite tuples, all required fields must satisfy the chosen matching criterion independently. A tuple is considered correct only when all required components match, and it is correctly attributed to its source paper.

\bmhead{Assignment of Predictions to Gold Tuples}
Multiple predictions may partially overlap with the same gold tuple, and multiple gold tuples may be similar within a paper. To avoid double-counting, matches are resolved under a one-to-one assignment constraint: each prediction can match at most one gold tuple, and each gold tuple can be matched by at most one prediction. Unmatched predictions reduce precision, while unmatched gold tuples reduce recall.

\bmhead{Multi-level matching}
We report results under three evaluation tiers that differ only in how strictly a predicted item must match the gold annotation:

\noindent - Strict (S): only exact matches are counted as correct.

\noindent - Medium (M): exact matches and over-detailed outputs are counted as correct.

\noindent - Lenient (L): exact matches, over-detailed outputs, and less-detailed outputs are counted as correct.

\noindent An over-detailed output preserves the correct core entity or relation while adding information beyond the gold annotation. A less detailed output preserves the correct core entity or relation while omitting some details present in the gold annotation.

\bmhead{Metrics}
For each paper and each query, we compute tuple-level precision, recall, and F1 over the set of predicted and gold tuples associated with that paper\rev{, following the standard definitions of these metrics. For tuple-level counting units}. A predicted tuple contributes to the set of correct matches only if it satisfies the selected matching criterion (Strict, Medium, or Lenient) and is correctly attributed to the originating paper.
Reported scores are \textbf{macro-averages}. We first compute precision, recall, and F1 independently for each paper, then average these paper-level scores across all papers within the corresponding evaluation setting. Precision, recall, and F1 are averaged separately rather than recomputed from aggregated counts.
This aggregation strategy gives equal weight to each paper regardless of how many tuples it contains, preventing overall scores from being dominated by a small number of tuple-rich documents.

\bmhead{Evaluation of Derived Statistical Queries}
The three output stages of computation-style queries are evaluated separately. (i) Evidences are scored using the same tuple-matching procedure as list-style extraction queries: precision, recall, and F1 are first computed within each paper, then macro-averaged across all papers in the evaluation set. (ii) Final list outputs are evaluated as intermediate structured results derived from contributing papers. Scores are computed at the relevant paper level and macro-averaged over papers that contribute gold-standard content to the queried result, so that non-contributing papers do not dilute the denominator. (iii) Final Answer is evaluated independently from paper identifiers and requires exact agreement with the value derived from gold annotations for scalar outputs (e.g., counts, mean, median), or exact set agreement for list-valued outputs.
Scoring these stages separately helps distinguish whether failures arise during evidence extraction, intermediate consolidation, or final answer generation.

\bmhead{LLM Judge for Tuple Matching}
All prediction--gold-matching decisions are adjudicated by an LLM judge under the three-level protocol (Strict, Medium, Lenient). The judge determines whether two tuples express the same underlying entity or relation while accounting for lexical variation, abbreviations, and differences in descriptive granularity that exact string matching cannot reliably capture. To verify that this adjudication procedure is trustworthy, we conducted an independent human validation study on a stratified sample of judged pairs. The validation results are reported below in this section.
Open-ended scientific extraction cannot be evaluated through exact string matching alone. Semantically equivalent study descriptors often appear in substantially different surface forms, including abbreviations, expanded names, regional variants, and differences in descriptive specificity. Examples in our corpora include ``USA'' vs.\ ``United States'', ``BMI'' vs.\ ``body mass index'', and ``children'' vs.\ ``children aged 5--18 years''. Conversely, high lexical overlap does not guarantee semantic identity, as superficially similar strings may denote different constructs. We therefore use an LLM judge to determine whether predicted and gold tuples express the same underlying entity or relation under the three-level protocol (Strict, Medium, Lenient).

\bmhead{Human Validation of the LLM Judge}
We stratified-sampled 100 prediction--gold pairs from the full pool of judge decisions, balanced across the four possible verdicts (\emph{exact}, \emph{over\_detail}, \emph{less\_detail}, \emph{no\_match}). Two external annotators with graduate training in computational linguistics, blind to the judge's decision and to each other, independently labeled all sampled pairs using the same scheme.
Human inter-annotator agreement was high (93.0\%; Cohen's $\kappa = 0.906$), \rev{indicating that the two independent annotators applied the four-way labeling scheme consistently, i.e., the matching criteria are well-defined enough to be reproduced across annotators}. On the subset with human consensus, the LLM judge agreed with the consensus label on 85\% of cases (Cohen's $\kappa = 0.80$), suggesting strong alignment with human semantic judgment. We therefore treat the judge as a sufficiently reliable proxy for tuple-level evaluation in this setting.
\rev{We note that this validation study assesses the reliability of the LLM judge as an evaluation instrument only. The two external annotators labeled prediction--gold match verdicts, not source articles, and their work therefore does not constitute an independent second annotation of the gold-standard data itself (see Section~\ref{sec:limitations}).}

\begin{table*}[!htb]
\begin{threeparttable}
\centering
\caption{Descriptive statistics of five domain-specific corpora, reporting summary statistics over counts of document-attributed schema atoms across domains. Token counts are computed on the \texttt{vlm}-parsed markdown of each paper using the \texttt{cl100k\_base} tokenizer.}
\small

\begin{tabular}{p{1.36cm} p{1.75cm}  p{1.7cm} p{1.7cm} p{1.7cm} p{1.7cm} p{1.55cm} p{0.95cm}}
\toprule
\textbf{Papers} & \textbf{Papers\tnote{a}} & \multicolumn{4}{l}{\textbf{Tokens per Paper}} & \textbf{\makecell{Unique}} & \textbf{\makecell{Effect}} \\
\cmidrule(lr){3-6}
\textbf{Counts} & \textbf{Source} & \textbf{Min} & \textbf{Median} & \textbf{Mean} & \textbf{Max} & \textbf{\makecell{Variables}} & \textbf{\makecell{Sizes}} \\
\toprule
\multicolumn{8}{l}{\textbf{Civil Engineering:} Building characteristics vs.\ Hotel energy use} \\
11 & \cite{arenhart2024energy} & 7{,}037 & 12{,}488 & 11{,}855 & 16{,}321 & 126 & 262 \\
\midrule
\multicolumn{8}{l}{\textbf{Medical and Health Science:} Anthropometric indicators vs.\ Ophthalmic outcomes} \\
 9 & \cite{Guan2025AnthropometricMyopia} & 8{,}319 & 10{,}967 & 11{,}873 & 17{,}375 & 49 & 328 \\
\midrule
\multicolumn{8}{l}{\textbf{Agricultural Science:} Environmental stress vs.\ Plant responses} \\
11 & Journal\tnote{b} & 10{,}936 & 15{,}888 & 15{,}125 & 18{,}355 & 40 & 236 \\
\midrule
\multicolumn{8}{l}{\textbf{Earth \& Environmental Sciences:} Carbon storage vs.\ Biodiversity metrics} \\
10 & \cite{liu2025exploring} & 8{,}285 & 14{,}253 & 15{,}894 & 26{,}042 & 78 & 199 \\
\midrule
\multicolumn{8}{l}{\textbf{Social Science:} Compassion vs.\ Well-being and burnout} \\
11 & \cite{Zhuniq2025well-being} & 9{,}108 & 15{,}794 & 15{,}986 & 28{,}642 & 62 & 153 \\
\bottomrule
\end{tabular}

\begin{tablenotes}
    \footnotesize
    \item[a] All papers are from meta-analysis review articles except specifically noted.
    \item[b] Retrieved \textit{Agricultural Water Management} (2005--2025) with query \textit{``(drought stress OR water stress) AND plant''}.
\end{tablenotes}

\label{tab:combined-corpus-stats}
\end{threeparttable}
\end{table*}

\section{Experiments}

This section describes the experimental setup, including the corpora, annotation protocol, preprocessing pipeline, model configurations, and input regimes. Evaluation follows the tuple-level metrics and procedures defined in Section~\ref{sec:method}.

\subsection{Corpora and Annotation}
\label{sec:corpora}
We construct five domain-specific corpora of primary empirical association studies spanning engineering, health, agricultural, environmental, and social sciences\rev{, comprising 52 full-text articles in total (9--11 per domain)}. The corpora are intentionally diverse in study scale, methodological complexity, and reporting practices, providing a challenging testbed for structured evidence extraction and derived statistical analysis. All included studies are primary empirical articles reporting quantitative associations using explicit statistical methods and effect size estimates; reviews and secondary analyses are excluded.
\rev{The corpus size is not arbitrary but follows from the design of the Global input regime, which requires all papers of a domain to be placed into a single prompt. Mainstream open-weight models currently provide context windows of roughly 196k--262k tokens, and full-text papers in our corpora average approximately 12k--16k tokens (Table~\ref{tab:combined-corpus-stats}). Around ten papers per domain therefore occupy a substantial fraction of the available window while leaving room for instructions and structured outputs. Five domains at this size yield the 52 papers used in this study.}
Table~\ref{tab:combined-corpus-stats} summarizes the five corpora. For each domain, it reports the corpus size along with summary statistics for document-attributed schema atoms, including sample sizes, statistical methods, variables, and reported effect-size instances. These statistics indicate substantial cross-domain variation in both study scale and the structural complexity of reported evidence.
Gold-standard annotations are created manually by a single expert annotator. All schema properties and association structures are extracted directly from the source text, preserving explicit document identifiers for every instance. Annotations include all reported variables, methods, and association estimates, regardless of statistical significance. Because the task emphasizes factual extraction, single-annotator curation was feasible; however, some normalization decisions remain judgment-sensitive. We therefore treat single-annotator annotation as a limitation of the study \rev{and discuss the associated risks in Section~\ref{sec:limitations}}.

\rev{\bmhead{Annotation protocol and codebook}
The annotator is the first author, a doctoral researcher in computer science specializing in scientific information extraction. What to annotate is fixed by the study-level schema in Table~\ref{tab:schema}, which was derived by abstracting recurring components from earlier ORKG domain templates (Section~\ref{sec:method}). The schema definitions were operationalized as a written codebook containing, for each field, a definition, inclusion and exclusion criteria, normalization rules (e.g., how to record sample sizes reported per subgroup, or units embedded in variable names), and worked examples. The codebook was developed in a pilot round in which a small number of papers from each domain were annotated first. Ambiguous cases that surfaced during the pilot, such as domain-specific reporting conventions for effect sizes, were resolved by discussion among the authors, and the resulting decision rules were added to the codebook before the main annotation pass.
To limit annotation fatigue, annotation proceeded in per-domain batches over several weeks rather than in a single continuous session, and every paper received a second verification pass in which the annotator re-checked all recorded tuples against the source text. Cross-disciplinary differences in reporting style are handled by the codebook's field-level decision rules rather than by per-domain schema changes, so that the same schema semantics apply uniformly across all five domains.}
We release all gold-standard JSON annotations, the full query suite, experiment scripts, and the DOIs of all included articles. Due to copyright restrictions, article PDFs and their derived text representations are not redistributed, but can be reconstructed from the released DOIs using the preprocessing pipeline described in Section~\ref{sec:setup}. This preserves methodological reproducibility while respecting publisher licenses.

\subsection{Modeling Conditions}
\label{sec:setup}

\bmhead{PDF-to-text preprocessing}

All source PDFs are converted to structured markdown using MinerU~2.5~\footnote{\url{https://github.com/opendatalab/MinerU}}~\cite{niu2025mineru25decoupledvisionlanguagemodel} with default settings. \rev{MinerU is an open-source tool that converts scholarly PDFs into machine-readable text, filling the same role as the OCR and text-extraction stage of a digital library ingestion pipeline. Rather than extracting a raw stream of characters, it reads the visual layout of each page and reconstructs the logical organization of the article.} The parser preserves document structure, reading order, tables, and mathematical expressions, producing markdown with HTML tables and \LaTeX formulas. No manual editing, segmentation, or truncation is applied after conversion. \rev{Individual papers average approximately 12--16k tokens (Table~\ref{tab:combined-corpus-stats}). For the Global regime, all papers within a domain are concatenated into a single input of approximately 107k--176k tokens, depending on the domain.}

\bmhead{Models}

We evaluate five LLMs spanning closed-source and open-weight families. 

\noindent (i) A closed-source frontier model:

\noindent - \texttt{OpenAI~GPT-5.4} (version 2026-03-05) \footnote{\url{https://platform.openai.com/docs/models/gpt-5.4}},

\noindent (ii) Four open-source models: 

\noindent - \texttt{google/gemma-4-31B-it} \footnote{\url{https://huggingface.co/google/gemma-4-31B-it}}, 

\noindent - \texttt{Qwen/Qwen3.5-122B-A10B-FP8} \footnote{\url{https://huggingface.co/Qwen/Qwen3.5-122B-A10B-FP8}}, 

\noindent - \texttt{Qwen/Qwen3.5-9B} \footnote{\url{https://huggingface.co/Qwen/Qwen3.5-9B}}, 

\noindent - \texttt{MiniMaxAI/MiniMax-M2.5} \footnote{\url{https://huggingface.co/MiniMaxAI/MiniMax-M2.5}}.

\noindent All models are used in text-only mode: inputs consist solely of MinerU markdown outputs, with no image content provided. Decoding uses temperature $=0.0$ and default settings otherwise. No retrieval tools or code execution are allowed, so results reflect intrinsic extraction and reasoning ability.
All domain inputs fit within each model's context window (Gemma~4~31B-it: 262k tokens; Qwen3.5-122B-A10B-FP8: 262k tokens; Qwen3.5-9B: 262k tokens; MiniMax-M2.5: 196k tokens; GPT-5.4: 1{,}050k tokens).
The four open-source models are served via vLLM~\cite{vllm} on $4\times$H100 GPUs\rev{. vLLM is a widely used open-source software package that loads the publicly released weights of such models and runs them on local computing hardware, much as an institution hosts its own repository or search services. All experiments with open-weight models therefore run entirely on local infrastructure and do not depend on an external provider}. GPT-5.4 is accessed via the OpenAI API.
To measure run-to-run stability, open-weight models are evaluated with three independent runs; GPT-5.4 is run once due to API cost constraints.

\rev{\bmhead{Rationale for evaluating models without agentic scaffolding}
We deliberately evaluate models in a bare configuration, without retrieval, tool use, or multi-agent orchestration, even though such frameworks are increasingly common in practice. Three considerations motivate this choice. First, agentic and retrieval-augmented systems are composed of the same underlying model calls, so the intrinsic reliability of schema-constrained extraction bounds and diagnoses the behavior of any framework built on top of it. Our results identify precisely the failure modes, such as multi-field binding and exact aggregation, that scaffolding must compensate for. Second, a scaffolding-free setup yields a reproducible, framework-agnostic baseline whose comparisons remain interpretable as orchestration strategies evolve. Third, our Per-paper regime already implements the most common orchestration primitive, namely task decomposition followed by aggregation, and the observed failures on exact corpus-level answers largely persist under it (Section~\ref{sec:results-pipeline}), indicating that these limitations are not removed by simple pipelining alone. The query suite itself is agnostic to the system under evaluation and can be applied unchanged to agentic pipelines in future work.}

\bmhead{Input Regimes and Query Execution}

The same tiered query suite is applied across all domains and models under two complementary input regimes: Global and Per-paper.
In the Global regime, all papers within a domain are concatenated into a single markdown input, and each corpus-level query is executed once in its native form. This setting directly evaluates long-context evidence extraction over multi-document inputs.
In the Per-paper regime, each article is processed independently before any cross-paper aggregation step. This provides a decomposition baseline that reduces context length and makes the provenance of extracted evidence explicit by construction.
Two considerations motivate this comparison. First, prior work reports position-dependent degradation in long-context models, where information distant from the query is used less effectively~\cite{liu2024LLMLostInLongContext}. Second, per-paper processing simplifies attribution because each extraction call sees only one source document.
Following prior work on decomposed and least-to-most prompting~\cite{zhou2023leasttomost,press-etal-2023-measuring,khot2023decomposed}, the Per-paper regime uses three stages: (i) rewrite the corpus-level query into a single-document version; (ii) execute it independently for each paper, without reusing intermediate outputs across papers; and (iii) aggregate the per-paper outputs in a final call using the same model.
List-style queries are rewritten by deterministic textual rules, while computation queries ($OC\_O1.\ast$ and $MC\_M\ast$ in Table~\ref{tab:computation-queries}) are rewritten by prompting the model to retain only the per-paper extraction sub-task and remove cross-paper aggregation steps.
Across all settings, the benchmark requires approximately 21,000 LLM calls \rev{ (20{,}956 in total, consisting of 1{,}690 Global calls, 17{,}576 Per-paper document-level calls, and 1{,}690 Per-paper aggregation calls, accumulated over all models, domains, queries, and runs)}.
Additional diagnostics on raw instruction-following and pre-rescue format compliance are reported in Appendix~\ref{secA1} and Table~\ref{tab:format-compliance}.
\rev{A detailed breakdown of computational cost and efficiency is provided in Appendix~\ref{sec:cost}.}

\begin{figure*}[t]
    \centering
    \includegraphics[width=0.99\linewidth]{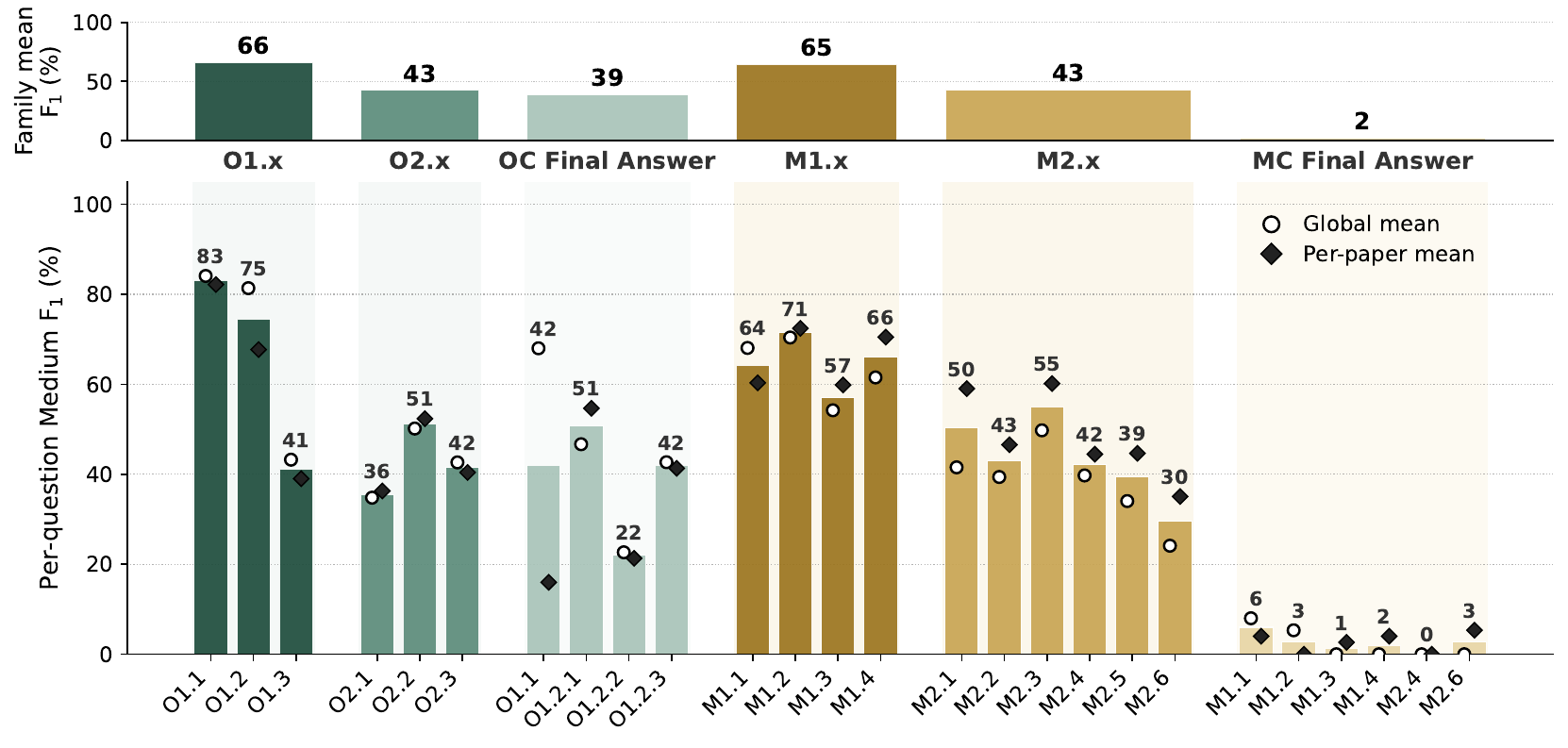}
    \caption{\textbf{Per-question performance profile across query suites}
(Medium F\textsubscript{1}, \%).
Bars show the five-model mean across Global and Per-paper settings for each question.
Circles and diamonds denote the corresponding Global and Per-paper means, respectively.
Top panels report family-level averages.
Performance generally declines from simpler extraction tasks (O1.x, M1.x) to more compositional settings (O2.x, M2.x), while Final Answer stages, especially MC, remain substantially harder than upstream evidence or list-construction steps.}
    \label{fig:pp_global_domain_drop}
\end{figure*}

\section{Results}
\label{sec:results}

We report findings along the structural progression defined in Section~\ref{sec:method}. The analysis proceeds from overall performance patterns (Section~\ref{sec:results-query-suites}), to precision--recall diagnostics (Section~\ref{sec:results-pr}), to sensitivity under alternative matching criteria (Section~\ref{sec:results-matching}), to pipeline effects (Section~\ref{sec:results-pipeline}), and finally to run-to-run stability (Section~\ref{sec:results-stability}). 
Unless otherwise noted, extraction tasks are reported as model-averaged paper-level macro F\textsubscript{1}, whereas Final Answer tasks use exact domain-level correctness.
Unless stated otherwise, extraction scores use the medium-matching tier, which counts exact and overly detailed predictions as correct.

\providecommand{\dA}[1]{#1}
\providecommand{\dB}[1]{\cellcolor{gd1}\textbf{#1}}
\providecommand{\dC}[1]{\cellcolor{gd2}\textbf{#1}}
\providecommand{\dD}[1]{\cellcolor{gd3}\textbf{#1}}


\begin{table}[t]
\centering
\caption{
\textbf{Effect of matching strictness on extraction performance}
(F\textsubscript{1}, \%, averaged across five models, five domains, and both Global / Per-paper pipelines).
\textbf{S} = strict (exact match only);
\textbf{M} = medium (exact + over-detail);
\textbf{L} = lenient (exact + over-detail + less-detail).
$\Delta$ = L $-$ S.
Shaded $\Delta$ cells indicate stronger sensitivity to matching strictness.
Relaxing match criteria substantially benefits open-ended extraction tasks (especially O1/O2), while final-answer tasks show little change.
}
\label{tab:strictness_final}

\small
\setlength{\tabcolsep}{6.7pt}
\renewcommand{\arraystretch}{1.16}

\begin{tabular}{
l c c c c
}
\toprule

\rowcolor{headbg}
\textbf{Question series}
& \textbf{S}
& \textbf{M}
& \textbf{L}
& $\boldsymbol{\Delta}$ \\

\midrule

O1.x
& 53.4 & 65.5 & 78.2 & \dD{24.8} \\

O2.x
& 20.7 & 42.7 & 71.3 & \dD{50.7} \\

\midrule

OC -- Evidences
& 71.6 & 75.9 & 77.0 & \dB{5.4} \\

OC -- Final List
& 78.3 & 80.3 & 80.3 & \dA{2.0} \\

OC -- Final Answer
& 39.0 & 39.0 & 39.0 & \dA{0.0} \\

\midrule

M1.x
& 56.1 & 64.0 & 66.9 & \dC{10.8} \\

M2.x
& 35.8 & 42.5 & 47.2 & \dC{11.4} \\

\midrule

MC -- Evidences
& 47.3 & 50.7 & 53.5 & \dB{6.2} \\

MC -- Final List
& 47.0 & 51.0 & 52.8 & \dB{5.8} \\

MC -- Final Answer
& 2.3 & 2.3 & 2.3 & \dA{0.0} \\

\bottomrule
\end{tabular}

\end{table}

\subsection{Performance Across Query Suites}

\label{sec:results-query-suites}

We first compare performance across query families and structural complexity levels (Table~\ref{tab:strictness_final}). For the extraction series, values are model-averaged paper-level macro F\textsubscript{1}; for the Final Answer rows, values reflect exact domain-level correctness. Because all models share the same query suites and scoring protocol, the comparison isolates the effect of structural demands on outputs.
Figure~\ref{fig:pp_global_domain_drop} shows the same pattern at per-question resolution: average performance is highest on simpler single-field tasks, declines on compositional extraction tasks, and drops most sharply on exact Final Answer stages. 
A full per-question tabular breakdown is provided in Appendix~\ref{secA2} and Table~\ref{tab:per-question}.

A clear hierarchy emerges. Single-field extraction scores highest: O1.x reaches 65.5 F\textsubscript{1} and M1.x reaches 64.0, showing that isolated object- and method-centric fields can be recovered with moderate reliability. Binding multiple fields into tuples causes a large drop: O2.x falls to 42.7 and the more relational M2.x to 42.5. Moving from \emph{identifying} fields to \emph{assembling} structured records is therefore a major source of error. The drop is sharper for downstream statistical queries. In the object-centric series, OC Evidences reaches 75.9, and OC Final List 80.3, but OC Final Answer falls to 39.0. The method-centric series shows the same pattern amplified: MC Evidences 50.7, MC Final List 51.0, yet MC Final Answer only 2.3. Models identify meaningful evidence and build partially correct candidate sets, but cannot deliver unique final values that require complete coverage, stable de-duplication, and exact aggregation. Object-centric tasks are consistently easier than their method-centric counterparts, and the gap widens with complexity. On direct extraction, O1.x slightly exceeds M1.x (65.5 vs.\ 64.0) and O2.x matches M2.x. On final answers, the contrast is much larger: 39.0 vs.\ 2.3. Counting countries or aggregating sample sizes is substantially easier than reasoning over variables, role assignments, relation pairs, and effect-size thresholds.
The results suggest that the key boundary is not simply between extraction and reasoning, but between locally plausible outputs and globally exact structured conclusions.


\definecolor{PrecA}{RGB}{240,235,247}
\definecolor{PrecB}{RGB}{223,212,238}
\definecolor{PrecC}{RGB}{199,183,225}
\definecolor{PrecD}{RGB}{171,149,207}

\definecolor{RecA}{RGB}{235,245,237}
\definecolor{RecB}{RGB}{210,232,214}
\definecolor{RecC}{RGB}{180,214,187}
\definecolor{RecD}{RGB}{143,190,154}

\definecolor{headbg}{RGB}{246,245,241}

\newcolumntype{C}[1]{>{\centering\arraybackslash}m{#1}}

\newcommand{\phA}[1]{\begingroup\setlength{\fboxsep}{0pt}\colorbox{PrecA}{\hspace{3pt}\strut #1\hspace{3pt}}\endgroup}
\newcommand{\phB}[1]{\begingroup\setlength{\fboxsep}{0pt}\colorbox{PrecB}{\hspace{3pt}\strut #1\hspace{3pt}}\endgroup}
\newcommand{\phC}[1]{\begingroup\setlength{\fboxsep}{0pt}\colorbox{PrecC}{\hspace{3pt}\strut #1\hspace{3pt}}\endgroup}
\newcommand{\phD}[1]{\begingroup\setlength{\fboxsep}{0pt}\colorbox{PrecD}{\hspace{3pt}\strut #1\hspace{3pt}}\endgroup}

\newcommand{\rhA}[1]{\begingroup\setlength{\fboxsep}{0pt}\colorbox{RecA}{\hspace{3pt}\strut #1\hspace{3pt}}\endgroup}
\newcommand{\rhB}[1]{\begingroup\setlength{\fboxsep}{0pt}\colorbox{RecB}{\hspace{3pt}\strut #1\hspace{3pt}}\endgroup}
\newcommand{\rhC}[1]{\begingroup\setlength{\fboxsep}{0pt}\colorbox{RecC}{\hspace{3pt}\strut #1\hspace{3pt}}\endgroup}
\newcommand{\rhD}[1]{\begingroup\setlength{\fboxsep}{0pt}\colorbox{RecD}{\hspace{3pt}\strut #1\hspace{3pt}}\endgroup}


\begin{table*}[t]
\centering
\caption{
\textbf{Precision--recall balance at medium matching tier}.
Global and Per-paper pipelines are shown separately.
Within each question block, shading compares Precision vs.\ Recall
for the same model and same pipeline:
\colorbox{PrecB}{purple} indicates Precision $>$ Recall, and
\colorbox{RecB}{green} indicates Recall $>$ Precision.
Darker shading indicates a larger gap.
}
\label{tab:pr_diag}

\scriptsize
\setlength{\tabcolsep}{2.2pt}
\renewcommand{\arraystretch}{1.12}

\begin{tabular}{
l p{2cm} p{1.6cm}
*{10}{C{0.93cm}}
}
\toprule

\rowcolor{headbg}
\multicolumn{3}{l}{\textbf{Question Series}}
& \multicolumn{2}{c}{GPT-5.4}
& \multicolumn{2}{c}{Gemma 4}
& \multicolumn{2}{c}{Qwen 3.5-122B}
& \multicolumn{2}{c}{MiniMax-M2.5}
& \multicolumn{2}{c}{Qwen 3.5-9B} \\

\cmidrule(lr){4-5}
\cmidrule(lr){6-7}
\cmidrule(lr){8-9}
\cmidrule(lr){10-11}
\cmidrule(lr){12-13}

& Step & Metric
& G & P & G & P & G & P & G & P & G & P \\

\midrule

O1.x & -- & Precision
& 70.9 & 67.7
& 68.2 & 68.4
& 76.0 & \phA{68.8}
& 66.2 & 42.8
& 64.9 & 66.5 \\

& & Recall
& \rhA{71.3} & 67.7
& 68.2 & 68.4
& \rhA{76.4} & 68.7
& \rhB{68.1} & \rhB{45.1}
& \rhC{68.1} & \rhA{66.9} \\

\cmidrule(lr){1-13}

O2.x & -- & Precision
& 45.5 & \phA{43.8}
& 33.6 & 38.9
& 46.9 & 53.0
& 39.6 & 40.6
& 46.0 & 37.8 \\

& & Recall
& \rhA{46.1} & 43.5
& 33.6 & 38.9
& 46.9 & 53.0
& \rhB{42.4} & \rhC{44.6}
& 46.0 & 37.8 \\

\midrule

OC & Evidences & Precision
& 81.7 & \phA{81.4}
& 81.0 & 86.0
& \phA{70.7} & \phA{81.2}
& 63.6 & \phA{72.2}
& \phA{67.6} & 78.0 \\

& & Recall
& 81.7 & 81.2
& 81.0 & 86.0
& 70.4 & 81.1
& \rhA{65.3} & \rhA{72.3}
& 67.2 & \rhB{79.0} \\

OC & Final List & Precision
& \phD{94.8} & \phD{91.8}
& \phD{95.4} & \phC{94.1}
& \phA{81.1} & \phA{85.6}
& \phC{68.0} & \phC{78.1}
& 73.7 & \phA{85.1} \\

& & Recall
& 86.9 & 80.3
& 81.1 & 87.3
& 78.7 & 85.0
& 61.7 & 72.6
& \rhA{73.9} & 84.1 \\

OC & Final Answer & Precision
& 60.0 & 20.0
& 45.0 & 30.0
& 53.3 & 51.7
& 26.7 & 30.0
& 40.0 & 35.0 \\

& & Recall
& 60.0 & 20.0
& 45.0 & 30.0
& 53.3 & 51.7
& 26.7 & 30.0
& 40.0 & 35.0 \\

\midrule

M1.x & -- & Precision
& \phC{69.9} & 70.7
& 71.6 & 70.3
& 66.4 & 68.9
& 61.8 & 65.1
& 57.7 & \phC{68.8} \\

& & Recall
& 78.6 & \rhB{73.7}
& \rhA{73.2} & \rhD{77.3}
& \rhC{71.9} & \rhA{69.6}
& \rhD{69.2} & \rhC{71.8}
& \rhD{64.7} & 64.6 \\

M2.x & -- & Precision
& \phA{50.1} & \phA{54.9}
& \phD{62.2} & \phA{60.6}
& 44.0 & \phD{55.9}
& \phC{38.2} & 44.2
& \phC{24.9} & \phD{57.1} \\

& & Recall
& 48.6 & 54.7
& 50.5 & 59.3
& \rhA{44.9} & 47.8
& 36.0 & \rhC{47.8}
& 20.2 & 39.6 \\

\midrule

MC & Evidences & Precision
& 55.1 & 64.7
& \phD{62.6} & 69.8
& \phA{44.5} & 52.7
& 47.2 & 59.9
& \phC{46.9} & \phC{57.7} \\

& & Recall
& \rhB{57.3} & \rhD{72.5}
& 52.8 & \rhA{71.7}
& 43.1 & \rhB{53.7}
& \rhA{48.7} & \rhD{64.0}
& 40.6 & 50.0 \\

MC & Final List & Precision
& 52.3 & 60.5
& \phD{69.7} & 65.6
& \phD{45.1} & \phC{61.7}
& \phD{50.4} & 60.5
& \phD{50.9} & \phD{61.0} \\

& & Recall
& \rhD{59.8} & \rhD{72.5}
& 49.6 & \rhD{71.2}
& 39.9 & 57.5
& 42.1 & \rhD{62.9}
& 39.1 & 51.2 \\

MC & Final Answer & Precision
& 0.0 & 6.7
& 2.2 & 5.6
& 4.4 & 1.1
& 1.1 & 0.0
& 3.3 & 0.0 \\

& & Recall
& 0.0 & 6.7
& 2.2 & 5.6
& 4.4 & 1.1
& 1.1 & 0.0
& 3.3 & 0.0 \\

\bottomrule
\end{tabular}
\end{table*}


\subsection{Precision and Recall}
\label{sec:results-pr}

Overall performance differences do not arise from a single source. To localize these failures, we first decompose medium-tier F\textsubscript{1} into precision and recall for both pipelines (Table~\ref{tab:pr_diag}). This analysis identifies whether tasks are limited mainly by over-generation, missing coverage, or both.

\bmhead{Balanced errors in object-centric}
For the simplest single-field tasks, precision and recall are usually close. In O1.x under the Global regime, several models remain near parity (GPT-5.4: 70.9 / 71.3; Gemma~4: 68.2 / 68.2; Qwen~3.5-122B: 76.0 / 76.4), and Per-paper shows a similar pattern for most systems. When only one schema field must be identified, models appear reasonably balanced between finding correct items and finding enough items. 
Errors at this stage appear more balanced than systematically one-sided.

\bmhead{Precision bottlenecks in method-centric}
M1.x shows a different but stable profile. Many models exhibit lower precision than recall, indicating over-generation. Under Global processing, this pattern appears consistently (GPT-5.4: 69.9 / 78.6; Qwen~3.5-122B: 66.4 / 71.9; MiniMax-M2.5: 61.8 / 69.2), and often persists under Per-paper. For method-related fields, models usually recover many plausible candidates, but add extra labels, boundary errors, or weakly grounded method mentions. 
The pattern suggests that candidate purity is often a larger issue than missing coverage.

\bmhead{Recall bottlenecks in set construction}
As tasks move from isolated fields to set construction, the dominant bottleneck often shifts to recall. The clearest example is OC Final List. Under Global processing, several models show high precision but noticeably lower recall (GPT-5.4: 94.8 / 86.9; Gemma~4: 95.4 / 81.1; MiniMax-M2.5: 68.0 / 61.7). The same tendency remains under Per-paper (Gemma~4: 94.1 / 87.3; MiniMax-M2.5: 78.1 / 72.6). These systems are usually accurate when they output an item, yet still fail to recover the full gold set. For object inventories, complete coverage appears harder than avoiding false positives.

\bmhead{Pipeline-induced error-mode reversal}
Method-centric relational tasks further reveal how the choice of pipeline affects system behavior. In M2.x, multiple models are recall-limited under Global settings (GPT-5.4: 50.1 / 48.6; Gemma~4: 62.2 / 50.5; Qwen~3.5-9B: 24.9 / 20.2). Per-paper often narrows this gap by improving recall (GPT-5.4: 54.9 / 54.7; Gemma~4: 60.6 / 59.3), suggesting that decomposition helps recover tuples that are harder to track in long shared contexts.
The clearest reversal appears in MC Final List. Under Global processing, most open-source models are recall-limited (Gemma~4: 69.7 / 49.6; Qwen~3.5-122B: 45.1 / 39.9; Qwen~3.5-9B: 50.9 / 39.1). Under Per-paper, several models flip to lower precision than recall (GPT-5.4: 60.5 / 72.5; Gemma~4: 65.6 / 71.2; MiniMax-M2.5: 60.5 / 62.9). Per-paper therefore recovers more target items, but also introduces more spurious candidates. It does not merely raise performance; it can change the dominant error mode from missing too much to reporting too much.
This also helps explain why gains on Final List do not consistently transfer to Final Answer. Missing items tend to underestimate counts or omit required relations, whereas redundant items inflate counts or contaminate candidate sets. Both regimes may improve intermediate F1, but their downstream numerical consequences differ.

\bmhead{Differences across model families}
Differences across model families are visible in balance, not only in score magnitude. GPT-5.4 usually maintains smaller precision--recall gaps, even on harder tasks, whereas several open-source models show stronger one-sided weaknesses in selected settings. For example, Qwen~3.5-9B on M2.x under Per-paper reaches 57.1 precision but only 39.6 recall, indicating that many required tuples are still missed despite reasonable candidate accuracy.

\bmhead{Summary}
Overall, the dominant error mode depends strongly on task structure. Simple field extraction is relatively balanced; method-field extraction more often overgenerates; object and relation sets more often underextract; and Per-paper frequently trades higher recall for lower precision on the hardest tasks. Progress in end-to-end scientific literature analytics will therefore require controlling coverage and purity jointly, rather than optimizing a single aggregate score.


\definecolor{GwinA}{RGB}{240,235,247}
\definecolor{GwinB}{RGB}{223,212,238}
\definecolor{GwinC}{RGB}{199,183,225}
\definecolor{GwinD}{RGB}{171,149,207}

\definecolor{PwinA}{RGB}{235,245,237}
\definecolor{PwinB}{RGB}{210,232,214}
\definecolor{PwinC}{RGB}{180,214,187}
\definecolor{PwinD}{RGB}{143,190,154}

\definecolor{headbg}{RGB}{246,245,241}

\newcolumntype{C}[1]{>{\centering\arraybackslash}m{#1}}

\newcommand{\gA}[1]{\begingroup\setlength{\fboxsep}{0pt}\colorbox{GwinA}{\hspace{2pt}\strut\textbf{#1}\hspace{2pt}}\endgroup}
\newcommand{\gB}[1]{\begingroup\setlength{\fboxsep}{0pt}\colorbox{GwinB}{\hspace{2pt}\strut\textbf{#1}\hspace{2pt}}\endgroup}
\newcommand{\gC}[1]{\begingroup\setlength{\fboxsep}{0pt}\colorbox{GwinC}{\hspace{2pt}\strut\textbf{#1}\hspace{2pt}}\endgroup}
\newcommand{\gD}[1]{\begingroup\setlength{\fboxsep}{0pt}\colorbox{GwinD}{\hspace{2pt}\strut\textbf{#1}\hspace{2pt}}\endgroup}

\newcommand{\pA}[1]{\begingroup\setlength{\fboxsep}{0pt}\colorbox{PwinA}{\hspace{2pt}\strut\textbf{#1}\hspace{2pt}}\endgroup}
\newcommand{\pB}[1]{\begingroup\setlength{\fboxsep}{0pt}\colorbox{PwinB}{\hspace{2pt}\strut\textbf{#1}\hspace{2pt}}\endgroup}
\newcommand{\pC}[1]{\begingroup\setlength{\fboxsep}{0pt}\colorbox{PwinC}{\hspace{2pt}\strut\textbf{#1}\hspace{2pt}}\endgroup}
\newcommand{\pD}[1]{\begingroup\setlength{\fboxsep}{0pt}\colorbox{PwinD}{\hspace{2pt}\strut\textbf{#1}\hspace{2pt}}\endgroup}


\begin{table*}[t]
\centering
\caption{
\textbf{Per-paper (P) vs.\ Global (G) pipelines across the full query suite}
(Medium F\textsubscript{1}, \%).
Each model column reports the series-level macro-average over
base units, domains, runs, and question identifiers within the series,
following Section~\ref{sec:evaluation-principles};
\emph{Overall} additionally macro-averages across the five models.
\colorbox{PwinB}{Green} indicates \textbf{G $<$ P};
\colorbox{GwinB}{purple} indicates \textbf{G $>$ P};
darker shading denotes larger margins.
Decomposed per-paper processing is most beneficial for structurally demanding extraction tasks, while global long-context prompting remains competitive for simpler extraction and some exact downstream answers.
}
\label{tab:main_results_final}

\vspace{2mm}
\scriptsize
\setlength{\tabcolsep}{2.2pt}
\renewcommand{\arraystretch}{1.12}

\begin{tabular}{
p{0.8cm} p{1.9cm}
*{12}{C{0.88cm}}
}

\toprule

\rowcolor{headbg}
\multicolumn{2}{l}{\textbf{Question Series}}
& \multicolumn{2}{c}{GPT-5.4}
& \multicolumn{2}{c}{Gemma 4}
& \multicolumn{2}{c}{Qwen 3.5-122B}
& \multicolumn{2}{c}{MiniMax-M2.5}
& \multicolumn{2}{c}{Qwen 3.5-9B}
& \multicolumn{2}{c}{\textbf{Overall}}
\\

\cmidrule(lr){3-4}
\cmidrule(lr){5-6}
\cmidrule(lr){7-8}
\cmidrule(lr){9-10}
\cmidrule(lr){11-12}
\cmidrule(lr){13-14}

& Step
& G & P & G & P & G & P & G & P & G & P & G & P \\

\midrule

O1.x & --
& \gA{71.0} & 67.7
& 68.2 & \pA{68.4}
& \gC{76.1} & 68.7
& \gD{66.8} & 43.4
& 65.7 & \pA{66.6}
& \gB{69.6} & 63.0 \\

O2.x & --
& \gA{45.7} & 43.6
& 33.6 & \pC{38.9}
& 46.9 & \pC{53.0}
& 40.4 & \pA{41.8}
& \gC{46.0} & 37.8
& 42.5 & \pA{43.0} \\

\midrule

OC & Evidences
& \gA{81.7} & 81.2
& 81.0 & \pC{86.0}
& 70.5 & \pD{81.1}
& 64.1 & \pD{72.2}
& 67.3 & \pD{78.3}
& 72.9 & \pD{79.8} \\

& Final List
& \gC{90.1} & 83.3
& 86.7 & \pA{89.5}
& 79.5 & \pC{84.9}
& 63.4 & \pD{73.9}
& 73.6 & \pD{84.0}
& 78.7 & \pA{83.1} \\

& Final Answer
& \gD{60.0} & 20.0
& \gC{45.0} & 30.0
& \gA{53.3} & 51.7
& 26.7 & \pA{30.0}
& \gA{40.0} & 35.0
& \gC{45.0} & 33.3 \\

\midrule

M1.x & --
& \gA{70.0} & 68.0
& 67.7 & \pA{69.7}
& 64.5 & \pA{64.7}
& 60.1 & \pA{64.6}
& 55.4 & \pC{61.9}
& 63.5 & \pA{65.8} \\

M2.x & --
& 44.7 & \pC{51.9}
& 51.1 & \pC{56.6}
& 40.6 & \pC{47.7}
& 33.8 & \pC{42.7}
& 20.3 & \pD{42.6}
& 38.1 & \pC{48.3} \\

\midrule

MC & Evidences
& 50.9 & \pD{64.5}
& 52.1 & \pD{67.0}
& 39.9 & \pC{48.6}
& 43.0 & \pD{57.6}
& 39.7 & \pC{49.4}
& 45.1 & \pD{57.4} \\

& Final List
& 52.6 & \pD{62.7}
& 54.6 & \pC{64.8}
& 39.1 & \pD{53.2}
& 42.0 & \pD{57.6}
& 39.9 & \pC{48.6}
& 45.6 & \pD{57.4} \\

& Final Answer
& 0.0 & \pC{6.7}
& 2.2 & \pA{5.6}
& \gA{4.4} & 1.1
& \gA{1.1} & 0.0
& \gA{3.3} & 0.0
& 2.2 & \pA{2.7} \\

\bottomrule
\end{tabular}
\end{table*}


\subsection{Matching Strictness}
\label{sec:results-matching}

We next examine how performance changes under different matching criteria: strict, medium, and lenient (Table~\ref{tab:strictness_final}). After identifying precision and recall bottlenecks, this comparison clarifies whether the remaining errors primarily reflect boundary or wording mismatches or substantive extraction failures.
Reported values remain averages across models, domains, and pipelines, with the underlying metric determined by task type. This comparison helps distinguish two sources of error. If scores rise sharply under looser criteria, many predictions preserve the correct semantic core but differ from the gold annotation. If scores remain unchanged, errors are more substantive.

The strongest gains appear in open-ended extraction tasks. For O2.x, average F1 rises from 20.7 under strict scoring to 42.7 under medium scoring, and further to 71.3 under lenient scoring. O1.x follows the same pattern, increasing from 53.4 to 65.5 and then to 78.2. These improvements indicate that many object-centric predictions are close to correct even when they fail exact matching. 
Models often appear to capture the target information but package it differently from the gold standard, using added detail, compressed phrasing, or alternative field boundaries.

Method-centric tasks improve more modestly. M1.x increases from 56.1 to 64.0 to 66.9, while M2.x rises from 35.8 to 42.5 to 47.2. Although looser matching recovers some credit, these scores remain below the object-centric series at comparable levels. Once tasks require variable roles, relation tuples, or coordinated multi-field outputs, performance limitations are not mainly due to wording mismatch.

Intermediate stages of the computational tasks show only limited sensitivity to the strictness of matching. OC Evidences increases from 71.6 to 75.9 to 77.0, and OC Final List changes only slightly from 78.3 to 80.3. MC Evidences rises from 47.3 to 50.7 to 53.5, while MC Final List increases from 47.0 to 51.0 to 52.8. This suggests that evidence sets and candidate lists are already closer to canonicalized outputs, leaving less room for score recovery through relaxed semantic matching.
By contrast, final statistical answers are completely unchanged. OC Final Answer remains 39.0 under all three criteria, and MC Final Answer remains 2.3 throughout. Because these outputs are exact counts, aggregates, or filtered sets rather than free-form spans, they do not benefit from tolerance in wording or detail level.

Overall, looser evaluation reveals that many upstream extraction errors are near misses rather than total failures. However, this flexibility largely disappears once the task requires a single correct analytical answer. Current models, therefore, appear closer to competent evidence retrieval than to reliable end-to-end structured reasoning.


\definecolor{stabA}{RGB}{236,244,238}
\definecolor{stabB}{RGB}{214,231,219}
\definecolor{stabC}{RGB}{248,242,223}
\definecolor{stabD}{RGB}{234,220,176}
\definecolor{stabE}{RGB}{219,210,235}

\definecolor{labelbg}{RGB}{250,251,248}
\definecolor{headbg}{RGB}{246,248,245}

\newcolumntype{L}[1]{>{\raggedright\arraybackslash}p{#1}}
\newcolumntype{C}[1]{>{\centering\arraybackslash}p{#1}}

\providecommand{\lab}[1]{\cellcolor{labelbg}#1}

\providecommand{\sA}[1]{%
\begingroup
\setlength{\fboxsep}{0pt}
\colorbox{stabA}{\hspace{2.8pt}\strut #1\hspace{2.8pt}}
\endgroup}

\providecommand{\sB}[1]{%
\begingroup
\setlength{\fboxsep}{0pt}
\colorbox{stabB}{\hspace{2.8pt}\strut #1\hspace{2.8pt}}
\endgroup}

\providecommand{\sC}[1]{%
\begingroup
\setlength{\fboxsep}{0pt}
\colorbox{stabC}{\hspace{2.8pt}\strut #1\hspace{2.8pt}}
\endgroup}

\providecommand{\sD}[1]{%
\begingroup
\setlength{\fboxsep}{0pt}
\colorbox{stabD}{\hspace{2.8pt}\strut\textbf{#1}\hspace{2.8pt}}
\endgroup}

\providecommand{\sE}[1]{%
\begingroup
\setlength{\fboxsep}{0pt}
\colorbox{stabE}{\hspace{2.8pt}\strut\textbf{#1}\hspace{2.8pt}}
\endgroup}


\begin{table*}[t]
\centering
\caption{
\textbf{Run-to-run stability of medium F\textsubscript{1}}
(standard deviation, percentage points \%) across three independent runs for open-source models.
Lower values indicate greater reproducibility.
Green cells denote stable settings; warmer cells indicate moderate variance;
purple cells indicate the highest volatility.
Most settings remain stable, while a few model--task combinations show substantial run-to-run variance.
}
\label{tab:stability}

\vspace{1mm}

\scriptsize
\setlength{\tabcolsep}{2.6pt}
\renewcommand{\arraystretch}{1.12}

\begin{tabular}{
L{1.2cm} L{2.1cm}
*{8}{C{1.35cm}}
}
\toprule

\rowcolor{headbg}

\multicolumn{2}{l}{\textbf{Question Series}} 
& \multicolumn{2}{c}{\makecell{\textbf{Gemma 4}\\(31B)}}
& \multicolumn{2}{c}{\makecell{\textbf{Qwen 3.5}\\(122B)}}
& \multicolumn{2}{c}{\makecell{\textbf{MiniMax-M2.5}\\(229B)}}
& \multicolumn{2}{c}{\makecell{\textbf{Qwen 3.5}\\(9B)}} \\

\cmidrule(lr){3-4}
\cmidrule(lr){5-6}
\cmidrule(lr){7-8}
\cmidrule(lr){9-10}

& Step 
& G & P & G & P & G & P & G & P \\

\midrule

\lab{O1.x} & --
& \sA{0.0} & \sB{1.0}
& \sA{0.9} & \sC{3.8}
& \sE{18.4} & \sD{6.1}
& \sB{1.4} & \sB{1.4} \\

\lab{O2.x} & --
& \sA{0.4} & \sA{0.0}
& \sB{1.5} & \sC{4.2}
& \sE{31.8} & \sE{18.3}
& \sA{0.0} & \sC{3.1} \\

\midrule

\lab{OC} & Evidences
& \sA{0.8} & \sA{0.0}
& \sB{1.6} & \sC{3.4}
& \sA{0.7} & \sD{6.4}
& \sA{0.0} & \sB{2.7} \\

& Final List
& \sA{0.8} & \sA{0.0}
& \sB{2.2} & \sB{2.8}
& \sB{2.3} & \sC{5.3}
& \sA{0.4} & \sA{0.4} \\

& Final Answer
& \sA{0.0} & \sA{0.0}
& \sE{7.6} & \sB{2.9}
& \sB{2.9} & \sC{5.0}
& \sA{0.0} & \sA{0.0} \\

\midrule

\lab{M1.x} & --
& \sB{1.6} & \sC{3.2}
& \sA{0.9} & \sC{3.3}
& \sB{2.1} & \sC{4.2}
& \sA{0.0} & \sB{2.9} \\

\lab{M2.x} & --
& \sA{0.6} & \sA{0.9}
& \sC{5.4} & \sB{1.4}
& \sB{1.7} & \sE{8.1}
& \sA{0.2} & \sA{0.3} \\

\midrule

\lab{MC} & Evidences
& \sA{0.2} & \sA{0.1}
& \sB{2.5} & \sB{1.8}
& \sA{0.6} & \sA{0.6}
& \sA{0.5} & \sA{0.1} \\

& Final List
& \sA{0.3} & \sA{0.0}
& \sB{1.4} & \sB{2.2}
& \sB{2.0} & \sB{1.3}
& \sA{0.5} & \sA{0.2} \\

& Final Answer
& \sB{1.9} & \sB{1.9}
& \sC{3.9} & \sB{1.9}
& \sB{1.9} & \sA{0.0}
& \sA{0.0} & \sA{0.0} \\

\midrule
\rowcolor{headbg}

\multicolumn{2}{l}{\textbf{Model Avg}}
& 0.7 & 0.7
& 2.8 & 2.8
& 6.4 & 5.5
& 0.3 & 1.1 \\

\bottomrule
\end{tabular}

\end{table*}



\subsection{Pipeline Comparison}
\label{sec:results-pipeline}

We next compare the two input regimes, Global and Per-paper, under the medium criterion (Table~\ref{tab:main_results_final}). Having identified the main error patterns above, we now test whether structured evidence extraction benefits more from shared long-context access or from document-level decomposition.

The overall pattern favors Per-paper on structurally complex tasks. In the multi-field method-centric series M2.x, average F1 rises from 38.1 under Global to 48.3 under Per-paper. Similar and even larger gains appear for MC Evidences (45.1 to 57.4) and MC Final List (45.6 to 57.4). Object-centric multi-step tasks show the same tendency: OC Evidences improves from 72.9 to 79.8, and OC Final List from 78.7 to 83.1. These tasks require aligning multiple fields within a study before any corpus-level synthesis occurs, so processing one paper at a time may better preserve local structural constraints.

Simpler extraction tasks show a different pattern. For O1.x, Global outperforms Per-paper overall (69.6 vs.\ 63.0). O2.x is nearly tied (42.5 vs.\ 43.0). When the target is a relatively direct field extraction rather than coordinated tuple recovery, the larger shared context of the Global regime may still provide useful cues without introducing serious interference.
However, stronger intermediate outputs do not automatically yield stronger final answers. Although Per-paper often improves Final List quality, the gains are much smaller or even reversed at the Final Answer stage. For object-centric computation, Global actually edges ahead of Per-paper on OC Final Answer (45.0 vs.\ 33.3), indicating that intermediate list gains do not necessarily propagate to the final aggregated statistic. For method-centric computation, both pipelines remain very low, with Global at 2.2 and Per-paper at 2.7. A candidate set can overlap well with the gold set while still containing enough omissions or redundancies to change the exact downstream statistic.

Pipeline effects also differ by model family. GPT-5.4 changes relatively little between regimes across most tasks, suggesting that this commercial frontier model is less sensitive to input organization in the present query suite. By contrast, several open-source models benefit substantially from decomposition. Qwen 3.5-9B improves on M2.x from 20.3 to 42.6, and Qwen 3.5-122B improves on OC Final List from 79.5 to 84.9. These results suggest that several open-source models benefit more from local document structure and may be more vulnerable to interference in large joint contexts.

Overall, the Per-paper pipeline is best understood not as a universal improvement, but as a targeted aid for complex structured extraction. It often raises the quality of intermediate evidence sets, especially for open-source models, yet it does not by itself solve the harder problem of producing exact corpus-level analytical answers.

\subsection{Run-to-Run Stability}
\label{sec:results-stability}

We finally examine run-to-run variance across three independent runs for the open-source models (Table~\ref{tab:stability}). Here the focus shifts from mean performance to reproducibility, reported as the standard deviation of medium-tier F\textsubscript{1}.

Overall, most model-task combinations show relatively low variance, indicating that the query suite produces stable performance signals. In many settings, standard deviation remains below three points, suggesting that the main comparisons reported above are not driven by one unusually strong or weak run.

Differences across model families are more pronounced than differences across task types. Gemma 4 and Qwen 3.5-9B are the most stable overall, with consistently low variance across both pipelines. Qwen 3.5-122B shows somewhat larger fluctuations, but remains broadly reliable. By contrast, MiniMax-M2.5 is substantially less stable than the other open-source models, with the highest average variance under both Global and Per-paper settings. This indicates that strong single-run performance does not necessarily imply reproducibility.
Instability is also not uniformly distributed across tasks. For MiniMax-M2.5, the largest fluctuations occur in O1.x and O2.x, where the standard deviations are 18.4 and 31.8 under Global processing. This is notable because these are comparatively simple extraction tasks. Variance, therefore, does not map directly onto task difficulty; it can also reflect model-specific generation behavior under certain output formats.
Pipeline choice affects consistency as well. For Qwen 3.5-9B, the average variance is somewhat lower under Global than under Per-paper processing, whereas Gemma 4 and Qwen 3.5-122B show nearly identical variance under both regimes. A plausible explanation is that the Per-paper regime introduces additional stages, including repeated document-level extraction and a separate aggregation step, each of which can add variability. The effect is visible for some but not all open-source models.

These findings indicate that stability is a distinct evaluation dimension alongside mean performance, particularly for scientific literature analytics where reproducibility matters.
A model that scores slightly lower but behaves consistently may be preferable to a stronger but volatile alternative, especially in scientific literature analytics where reproducibility is essential.

\section{\rev{Limitations}}
\label{sec:limitations}

\rev{\bmhead{Single-annotator gold standard}
The gold annotations were produced by one expert annotator following the written codebook described in Section~\ref{sec:setup}. Although the task is predominantly factual extraction, and every paper received a second verification pass, a single-annotator design carries two risks that we cannot fully rule out, namely annotation fatigue over a large number of tuples and idiosyncratic normalization choices where reporting conventions differ across disciplines. The human validation study in Section~\ref{sec:evaluation-principles} demonstrates that the matching criteria can be applied consistently by independent annotators, but it validates the LLM judge, not the gold data, and it is not a substitute for independent dual annotation of the corpus. Absolute scores should therefore be read with this caveat in mind, whereas the relative comparisons across models, task tiers, and input regimes, which share the same gold standard, are less affected by any systematic annotator bias.}

\rev{\bmhead{Corpus size}
The five corpora comprise 52 papers. As explained in Section~\ref{sec:corpora}, this size follows from the requirement that each domain fit into the context window of current open-weight models under the Global regime, but it remains modest. The study is accordingly positioned as a diagnostic evaluation of failure modes rather than an exhaustive benchmark. Extending the framework to larger corpora under the Per-paper regime is straightforward and is left to future work.}

\rev{\bmhead{Scope of the evaluated systems}
We evaluate bare models without retrieval, tools, or agentic orchestration, for the reasons given in Section~\ref{sec:setup}. Results should therefore be interpreted as a lower bound on what composite systems may achieve, and as a diagnosis of the component-level weaknesses such systems must address. In addition, GPT-5.4 was run once due to API cost, so its run-to-run variance is not characterized.}

\section{Conclusion}

We presented a structural and diagnostic evaluation of LLM-based scientific evidence extraction, treating the task as a progression from single-field recovery to multi-field relational binding and corpus-level analytical derivation under a shared study schema. Across five scientific domains, five contemporary language models, and two input regimes, the results reveal a persistent gap between locally plausible extraction outputs and globally reliable structured evidence use.

Current models are substantially stronger at recovering isolated attributes than at preserving cross-field dependencies required for coherent tuples and exact downstream answers. Small upstream omissions, duplications, or misbindings often become decisive once tasks require exact counts, filtered sets, or aggregate statistics. This suggests that evidence extraction errors are not merely additive; they are amplified under strict downstream composition.

Our comparison of global long-context prompting and per-paper decomposition further clarifies these limitations. Processing one article at a time often improves intermediate extraction quality, especially on structurally complex tasks, but these gains transfer only weakly to final analytical correctness. Therefore, improved retrieval of local evidence is insufficient without mechanisms that enforce consistency during aggregation.

Overall, structured scientific evidence extraction appears less like conventional span extraction and more like a constrained reasoning problem over linked study records. The query suite and staged evaluation protocol introduced here provide a reusable framework for diagnosing such failures and benchmarking future systems. Promising directions include schema-constrained decoding, explicit symbolic intermediate representations, verifier modules, and hybrid pipelines that separate extraction from deterministic aggregation. Advancing these capabilities is essential for automated literature-analysis systems intended for evidence synthesis, systematic review, and meta-analysis.

\backmatter


\section*{Declarations}

\bmhead{Funding}
This work was supported by the project ``HybrInt -- Hybrid Intelligence through Interpretable AI in Machine Perception and Interaction'' (Zukunft.Niedersachsen; Lower Saxony Ministry for Science and Culture; Grant ID: ZN4219). Computational resources were provided by the TIB -- Leibniz Information Centre for Science and Technology.

\bmhead{Conflict of interest}
The authors declare that they have no competing interests, financial or non-financial, that could have influenced the work reported in this article.

\bmhead{Ethics approval and consent to participate}
Not applicable. This study does not involve human participants, human data, or animal subjects. The analyzed materials consist entirely of previously published scientific articles retrieved from publishers and open repositories, used in accordance with their respective licenses.

\bmhead{Consent for publication}
Not applicable. No identifiable personal data or images are reported.

\bmhead{Data availability}
The gold-standard JSON annotations, the full tiered query suite, the prompts used for extraction and for the LLM judge, and the DOIs of all articles included in the five domain corpora are available in the project repository at \url{https://github.com/zhiyintan/llm-scientific-evidence-extraction}. Due to copyright restrictions, the source article PDFs and their MinerU-derived markdown representations are not redistributed. They can be reconstructed from the released DOIs using the preprocessing pipeline described in Section~\ref{sec:setup}.

\bmhead{Materials availability}
Not applicable. This study does not involve physical materials.

\bmhead{Code availability}
All code required to reproduce the experiments and analyses reported in this paper, including the extraction pipeline for both global and per-paper input regimes, the evaluation pipeline with tuple-level matching and LLM-judge adjudication, and the scripts that generate all tables and figures, is publicly available in the project repository at \url{https://github.com/zhiyintan/llm-scientific-evidence-extraction} under an open-source license.

\bmhead{Author contribution}
Author\textsuperscript{1} designed the semantic study schema and the tiered query suite, implemented the extraction and evaluation pipelines, conducted the experiments, performed the analyses, and drafted the manuscript. Authors\textsuperscript{2,3} critically discussed and revised the manuscript. All authors read and approved the final version of the manuscript.

\noindent

\begin{appendices}

\section{\rev{Cost and Efficiency}}
\label{sec:cost}

\rev{Because the benchmark evaluates five models across two input regimes, five domains, and repeated runs for open-weight systems, computational cost is relevant for reproducibility and future scaling. This section, moved from the main text, details the request volume and runtime.}

\bmhead{Extraction}
A full extraction pass issues \rev{20{,}956 LLM calls in total. With 26 queries per domain and five domains, the Global regime issues $26 \times 5 = 130$ calls per model run, i.e., $130 \times 13 = 1{,}690$ calls across all runs (three runs for each of the four open-weight models plus one run for GPT-5.4). The Per-paper regime issues one document-level call per query and paper ($26 \times 52 = 1{,}352$ per run and $17{,}576$ across all runs) followed by one aggregation call per query and domain ($1{,}690$ across all runs).} The Per-paper regime requires more calls than the Global regime because each corpus-level query is decomposed into document-level executions followed by a final aggregation step. However, each call uses substantially shorter inputs, so wall-clock cost remains comparable between regimes.

\bmhead{Evaluation}
A complete scoring pass over all outputs, including tuple matching, LLM-judge adjudication for ambiguous pairs, and the three-stage computation-task pipeline, requires approximately four hours. Most evaluation cost arises from LLM-based judging; programmatic scoring contributes negligibly.

\bmhead{Scalability}
Under the hardware configuration described above, full reproduction requires roughly 48 hours end-to-end. Scaling to larger corpora would primarily increase Per-paper extraction cost approximately linearly with document count, whereas the Global regime is constrained mainly by model context limits.

\section{Instruction-following and format compliance}\label{secA1}

The main-text result tables report scores computed on predictions after our deterministic re-parsing and LLM-based rescue stage.
To characterize the intrinsic instruction-following
behavior of each model, Table~\ref{tab:format-compliance} reports the raw
format-compliance rate measured on the pre-rescue predictions.

\begin{table*}[htbp]
\centering
\caption{%
  Instruction-following / format compliance per model and pipeline,
  measured on the pre-rescue predictions (i.e., before any deterministic
  re-parsing or LLM-based rescue is applied).
  \emph{err rate} = fraction of (paper, question) entries whose raw output is
  not a well-formed, non-empty list under the expected schema key
  (\texttt{answers} for list-style queries, \texttt{evidences} for
  computation queries). Reasons: \texttt{wrong key} dict present but
  uses an incorrect key; \texttt{empty list} correct key but empty;
  \texttt{non json} type mismatch. $n$ denotes the
  total number of (paper,~question) entries scanned per cell (aggregated
  across all three runs for open-source models, one run for GPT-5.4).
}
\label{tab:format-compliance}
\setlength{\tabcolsep}{9pt}
\small
\begin{tabular}{l c r r r r r}
\toprule
\textbf{Model} & \textbf{Mode} & \textbf{error rate} & \textbf{wrong key} & \textbf{empty list} & \textbf{non json} & \textbf{$n$} \\
\midrule
GPT-5.4                       & Global &  2.3\% &   3 &   0 & 0 & 130 \\
GPT-5.4                       & Per-paper &  5.4\% &   2 &   5 & 0 & 130 \\
Gemma\,4 (31B)                & Global &  0.0\% &   0 &   0 & 0 & 390 \\
Gemma\,4 (31B)                & Per-paper &  0.0\% &   0 &   0 & 0 & 390 \\
Qwen\,3.5 (122B)              & Global & 10.3\% &   1 &  39 & 0 & 390 \\
Qwen\,3.5 (122B)              & Per-paper & 41.5\% & 162 &   0 & 0 & 390 \\
MiniMax-M2.5 (229B)           & Global &  2.1\% &   1 &   7 & 0 & 390 \\
MiniMax-M2.5 (229B)           & Per-paper &  3.1\% &  10 &   0 & 2 & 390 \\
Qwen\,3.5 (9B)                & Global & 16.9\% &   0 &  66 & 0 & 390 \\
Qwen\,3.5 (9B)                & Per-paper & 65.4\% & 240 &  15 & 0 & 390 \\
\bottomrule
\end{tabular}
\end{table*}

The dominant failure mode under the \emph{per\_paper} regime is
\texttt{wrong key}: the model returns a top-level dictionary whose key is
not the one requested by the prompt (e.g., \texttt{"result"} or
\texttt{"data"} instead of \texttt{"answers"}). Under \emph{global}, the
dominant mode is \texttt{empty list}: the correct key is emitted, but the
associated list is empty, typically on the harder MC/OC final-list steps.
Format compliance scales sharply with model size within a family: the 9B
Qwen variant exhibits 65\% broken per-paper outputs, whereas the 122B
variant exhibits 42\%; Gemma\,4 (31B) is fully compliant. This pattern is
orthogonal to overall F\textsubscript{1} quality, and motivates the rescue
stage used before scoring.


\section{Per-question F\textsubscript{1} breakdown}\label{secA2}

Tables~\ref{tab:main_results_final} and~\ref{tab:strictness_final} in
the main text aggregate results by question \emph{series} (O1.x, O2.x,
M1.x, M2.x). Table~\ref{tab:per-question} below expands these series
into their individual list-style questions, using the same
\textbf{Global} vs.\ \textbf{Per-paper} comparison style.

\definecolor{psoft}{RGB}{232,244,234}
\definecolor{pstrong}{RGB}{206,232,210}
\definecolor{labelbg}{RGB}{250,251,248}
\providecommand{\pw}[1]{\cellcolor{psoft}\textbf{#1}}
\providecommand{\ps}[1]{\cellcolor{pstrong}\textbf{#1}}
\providecommand{\best}[1]{\textbf{#1}}
\providecommand{\lab}[1]{\cellcolor{labelbg}#1}

\begin{table*}[!htbp]
\centering
\caption{
\textbf{Per-question Per-paper (P) vs.\ Global (G) breakdown}
(Medium F\textsubscript{1}, \%), expanded from the aggregated series rows of
Table~\ref{tab:main_results_final}.
Green-shaded cells indicate settings where \textbf{P outperforms G};
darker shading denotes larger gains ($\geq$5 points).
Each cell is the mean across runs (three independent runs for the four
open-source models, one run for GPT-5.4) and the five domains.
}
\label{tab:per-question}

\small
\setlength{\tabcolsep}{4.8pt}

\begin{tabular}{
L{3.7cm}
*{10}{C{0.85cm}}
}
\toprule

& \multicolumn{2}{c}{\shortstack{\textbf{GPT-5.4}}}
& \multicolumn{2}{c}{\shortstack{\textbf{Gemma 4}\\(31B)}}
& \multicolumn{2}{c}{\shortstack{\textbf{Qwen 3.5}\\(122B)}}
& \multicolumn{2}{c}{\shortstack{\textbf{MiniMax-}\\\textbf{M2.5}(229B)}}
& \multicolumn{2}{c}{\shortstack{\textbf{Qwen 3.5}\\(9B)}}
\\

\cmidrule(lr){2-3}
\cmidrule(lr){4-5}
\cmidrule(lr){6-7}
\cmidrule(lr){8-9}
\cmidrule(lr){10-11}

\textbf{Question}
& G & P & G & P & G & P & G & P & G & P \\

\midrule

\lab{O1.1}
& 86.0 & \ps{92.7}
& 87.1 & \pw{90.3}
& \best{91.3} & 75.1
& \best{75.8} & 64.2
& 80.2 & \ps{88.5} \\

\lab{O1.2}
& \best{86.7} & 77.6
& 83.1 & \pw{86.5}
& \best{85.6} & 75.4
& \best{78.9} & 18.4
& 72.5 & \ps{80.7} \\

\lab{O1.3}
& \best{40.4} & 32.7
& \best{34.5} & 28.5
& 51.3 & \pw{55.8}
& 45.6 & \pw{47.4}
& \best{44.4} & 30.7 \\

\midrule

\lab{O2.1}
& 34.7 & \pw{36.4}
& 25.5 & \ps{36.4}
& 39.5 & \ps{49.1}
& \best{42.7} & 29.1
& \best{31.6} & 30.5 \\

\lab{O2.2}
& \best{60.0} & 58.5
& 40.0 & \ps{45.0}
& 49.9 & \ps{63.0}
& 45.7 & \ps{53.3}
& \best{55.2} & 42.0 \\

\lab{O2.3}
& \best{42.4} & 35.9
& \best{35.5} & 35.3
& \best{51.3} & 46.9
& 32.9 & \ps{43.0}
& \best{51.1} & 40.9 \\

\midrule
\lab{OC\_O1.1/Evidences} & 73.6 & \ps{92.1} & 88.1 & \pw{88.9} & 68.0 & \ps{77.3} & \best{85.7} & 72.5 & 65.6 & \ps{76.6} \\
\lab{OC\_O1.1/Final List} & \best{92.5} & 75.5 & \best{92.8} & 83.4 & \best{93.2} & 78.5 & \best{98.5} & 73.5 & \best{90.4} & 87.0 \\
\lab{OC\_O1.1/Final Answer} & \best{60.0} & 0.0 & \best{60.0} & 0.0 & \best{60.0} & 13.3 & \best{80.0} & 26.7 & \best{80.0} & 40.0 \\
\lab{OC\_O1.2.1/Evidences} & \best{81.3} & 77.6 & 78.4 & \ps{85.1} & 74.9 & \ps{82.4} & 60.7 & \ps{72.0} & \best{86.3} & 78.9 \\
\lab{OC\_O1.2.1/Final List} & \best{89.8} & 88.2 & 84.4 & \ps{93.5} & 81.8 & \ps{89.7} & 40.6 & \ps{71.9} & \best{88.7} & 86.8 \\
\lab{OC\_O1.2.1/Final Answer} & \best{60.0} & 40.0 & \best{40.0} & \best{40.0} & 66.7 & \ps{80.0} & 6.7 & \ps{53.3} & \best{60.0} & \best{60.0} \\
\lab{OC\_O1.2.2/Evidences} & \best{84.9} & 77.6 & 77.1 & \ps{85.1} & 61.4 & \ps{82.4} & 61.4 & \ps{72.0} & 57.8 & \ps{78.9} \\
\lab{OC\_O1.2.2/Final List} & \best{89.1} & 84.8 & 83.4 & \ps{90.7} & 62.4 & \ps{85.6} & 64.4 & \ps{75.1} & 57.8 & \ps{81.1} \\
\lab{OC\_O1.2.2/Final Answer} & \best{60.0} & 20.0 & 20.0 & \ps{40.0} & 20.0 & \ps{33.3} & \best{13.3} & \best{13.3} & \best{0.0} & \best{0.0} \\
\lab{OC\_O1.2.3/Evidences} & \best{86.9} & 77.6 & 80.4 & \pw{85.1} & 77.7 & \pw{82.4} & 48.7 & \ps{72.0} & 59.6 & \ps{78.9} \\
\lab{OC\_O1.2.3/Final List} & \best{89.1} & 84.8 & 86.1 & \pw{90.7} & 80.5 & \ps{85.6} & 50.4 & \ps{75.1} & 57.8 & \ps{81.1} \\
\lab{OC\_O1.2.3/Final Answer} & \best{60.0} & 20.0 & \best{60.0} & 40.0 & 66.7 & \ps{80.0} & 6.7 & \ps{26.7} & 20.0 & \ps{40.0} \\

\midrule
\lab{M1.1}
& \best{72.7} & 61.9
& \best{74.4} & 62.3
& \best{73.7} & 61.3
& 59.5 & \pw{62.3}
& \best{60.1} & 53.8 \\

\lab{M1.2}
& 77.5 & \pw{77.6}
& 71.8 & \ps{79.2}
& 70.1 & \pw{71.7}
& 67.3 & \pw{67.6}
& 65.3 & \pw{66.0} \\

\lab{M1.3}
& \best{63.0} & 60.7
& 55.4 & \ps{60.9}
& 48.2 & \ps{61.4}
& 54.6 & \pw{56.5}
& 49.9 & \ps{59.8} \\

\lab{M1.4}
& 66.9 & \pw{71.9}
& 69.1 & \ps{76.3}
& \best{66.0} & 64.3
& 59.1 & \ps{72.0}
& 46.5 & \ps{67.9} \\

\midrule

\lab{M2.1}
& 55.7 & \ps{66.1}
& 57.1 & \pw{60.9}
& 40.3 & \ps{59.1}
& 54.6 & \pw{57.4}
& 0.0 & \ps{51.6} \\

\lab{M2.2}
& \best{46.4} & 44.9
& 47.1 & \ps{54.5}
& 35.0 & \ps{53.0}
& \best{40.7} & 30.1
& 27.8 & \ps{50.4} \\

\lab{M2.3}
& 58.4 & \ps{68.0}
& \best{63.1} & \best{63.1}
& 42.9 & \ps{59.7}
& \best{58.4} & 53.4
& 26.0 & \ps{56.6} \\

\lab{M2.4}
& \best{40.9} & 40.0
& 58.3 & \pw{58.9}
& \best{48.5} & 48.4
& 20.8 & \ps{31.0}
& 30.2 & \ps{44.0} \\

\lab{M2.5}
& 44.0 & \pw{47.9}
& 40.1 & \ps{56.4}
& 32.8 & \ps{42.0}
& 22.9 & \ps{42.3}
& 30.4 & \pw{34.7} \\

\lab{M2.6}
& 22.8 & \ps{44.7}
& 40.7 & \ps{46.1}
& \best{44.1} & 24.0
& 5.3 & \ps{42.3}
& 7.7 & \ps{18.1} \\

\midrule
\lab{MC\_M1.1/Evidences} & 62.6 & \ps{73.0} & 69.1 & \pw{69.6} & 58.7 & \pw{61.5} & 60.7 & \ps{68.8} & 48.5 & \ps{65.1} \\
\lab{MC\_M1.1/Final List} & 51.3 & \ps{60.4} & 44.8 & \ps{60.4} & 48.4 & \ps{57.2} & 43.3 & \ps{59.5} & 48.1 & \ps{57.0} \\
\lab{MC\_M1.1/Final Answer} & 0.0 & \ps{20.0} & \best{0.0} & \best{0.0} & \best{13.3} & 0.0 & \best{6.7} & 0.0 & \best{20.0} & 0.0 \\
\lab{MC\_M1.2/Evidences} & 74.5 & \ps{79.6} & 71.0 & \ps{79.0} & 43.9 & \ps{65.1} & 60.8 & \ps{70.0} & 59.5 & \ps{65.2} \\
\lab{MC\_M1.2/Final List} & 73.0 & \pw{77.4} & 70.7 & \ps{76.5} & 42.8 & \ps{64.1} & 61.1 & \ps{67.7} & 54.7 & \ps{63.1} \\
\lab{MC\_M1.2/Final Answer} & \best{0.0} & \best{0.0} & \best{13.3} & 0.0 & \best{13.3} & 0.0 & \best{0.0} & \best{0.0} & \best{0.0} & \best{0.0} \\
\lab{MC\_M1.3/Evidences} & 59.5 & \ps{66.6} & 50.1 & \ps{66.8} & 40.0 & \ps{56.4} & 46.3 & \ps{52.1} & 41.9 & \ps{55.3} \\
\lab{MC\_M1.3/Final List} & 60.5 & \ps{67.8} & 55.3 & \ps{62.6} & 48.2 & \ps{60.0} & 45.8 & \ps{56.8} & 47.5 & \ps{53.4} \\
\lab{MC\_M1.3/Final Answer} & \best{0.0} & \best{0.0} & 0.0 & \ps{13.3} & \best{0.0} & \best{0.0} & \best{0.0} & \best{0.0} & \best{0.0} & \best{0.0} \\
\lab{MC\_M1.4/Evidences} & 61.9 & \ps{71.0} & 65.1 & \ps{75.3} & 53.9 & \ps{62.2} & 53.1 & \ps{59.2} & 57.2 & \ps{66.6} \\
\lab{MC\_M1.4/Final List} & 64.5 & \ps{70.8} & 67.7 & \ps{75.2} & 56.3 & \ps{63.1} & 55.8 & \ps{63.1} & \best{62.6} & 62.5 \\
\lab{MC\_M1.4/Final Answer} & 0.0 & \ps{20.0} & \best{0.0} & \best{0.0} & \best{0.0} & \best{0.0} & \best{0.0} & \best{0.0} & \best{0.0} & \best{0.0} \\
\lab{MC\_M2.4/Evidences} & 40.4 & \ps{49.7} & 48.0 & \ps{61.1} & \best{40.4} & 40.0 & 31.8 & \ps{50.9} & 30.8 & \ps{42.1} \\
\lab{MC\_M2.4/Final List} & 34.3 & \ps{51.3} & 46.4 & \ps{63.7} & 34.8 & \ps{45.3} & 32.3 & \ps{54.3} & 26.4 & \ps{39.9} \\
\lab{MC\_M2.4/Final Answer} & \best{0.0} & \best{0.0} & \best{0.0} & \best{0.0} & \best{0.0} & \best{0.0} & \best{0.0} & \best{0.0} & \best{0.0} & \best{0.0} \\
\lab{MC\_M2.6/Evidences} & 6.8 & \ps{47.3} & 9.0 & \ps{50.6} & 2.2 & \pw{6.3} & 5.2 & \ps{44.5} & \best{0.0} & \pw{2.1} \\
\lab{MC\_M2.6/Final List} & 32.0 & \ps{48.5} & 42.7 & \ps{50.4} & 4.4 & \ps{29.7} & 13.8 & \ps{44.1} & \best{0.0} & 15.6 \\
\lab{MC\_M2.6/Final Answer} & \best{0.0} & \best{0.0} & 0.0 & \ps{20.0} & 0.0 & \ps{6.7} & \best{0.0} & \best{0.0} & \best{0.0} & \best{0.0} \\

\bottomrule
\end{tabular}

\end{table*}

\clearpage
\end{appendices}

\bibliography{sn-bibliography}

\end{document}